
\hsize=6.5truein
\hoffset=.3truein
\vsize=8.9truein
\voffset=.1truein
\font\twelverm=cmr10 scaled 1200    \font\twelvei=cmmi10 scaled 1200
\font\twelvesy=cmsy10 scaled 1200   \font\twelveex=cmex10 scaled 1200
\font\twelvebf=cmbx10 scaled 1200   \font\twelvesl=cmsl10 scaled 1200
\font\twelvett=cmtt10 scaled 1200   \font\twelveit=cmti10 scaled 1200
\skewchar\twelvei='177   \skewchar\twelvesy='60
\def\twelvepoint{\normalbaselineskip=14pt
  \abovedisplayskip 12.4pt plus 3pt minus 9pt
  \belowdisplayskip 12.4pt plus 3pt minus 9pt
  \abovedisplayshortskip 0pt plus 3pt
  \belowdisplayshortskip 7.2pt plus 3pt minus 4pt
  \smallskipamount=3.6pt plus1.2pt minus1.2pt
  \medskipamount=7.2pt plus2.4pt minus2.4pt
  \bigskipamount=14.4pt plus4.8pt minus4.8pt
  \def\rm{\fam0\twelverm}          \def\it{\fam\itfam\twelveit}%
  \def\sl{\fam\slfam\twelvesl}     \def\bf{\fam\bffam\twelvebf}%
  \def\mit{\fam 1}                 \def\cal{\fam 2}%
  \def\tt{\twelvett}
  \textfont0=\twelverm   \scriptfont0=\tenrm   \scriptscriptfont0=\sevenrm
  \textfont1=\twelvei    \scriptfont1=\teni    \scriptscriptfont1=\seveni
  \textfont2=\twelvesy   \scriptfont2=\tensy   \scriptscriptfont2=\sevensy
  \textfont3=\twelveex   \scriptfont3=\twelveex  \scriptscriptfont3=\twelveex
  \textfont\itfam=\twelveit
  \textfont\slfam=\twelvesl
  \textfont\bffam=\twelvebf \scriptfont\bffam=\tenbf
  \scriptscriptfont\bffam=\sevenbf
  \normalbaselines\rm}

\def\beginlinemode{\endmode
  \begingroup\parskip=0pt \obeylines\def\\{\par}\def\endmode{\par\endgroup}}
\def\beginparmode{\endmode
  \begingroup \def\endmode{\par\endgroup}}
\let\endmode=\par
{\obeylines\gdef\
{}}
\def\singlespace{\baselineskip=\normalbaselineskip}
\def\oneandahalfspace{\baselineskip=\normalbaselineskip
  \multiply\baselineskip by 3 \divide\baselineskip by 2}
\def\doublespace{\baselineskip=\normalbaselineskip \multiply\baselineskip by 2}
\newcount\firstpageno
\firstpageno=2
\footline={\ifnum\pageno<\firstpageno{\hfil}\else{\hfil\twelverm\folio\hfil}\fi}
\let\rawfootnote=\footnote              
\def\footnote#1#2{{\rm\singlespace\parindent=0pt\rawfootnote{#1}{#2}}}
\def\raggedcenter{\leftskip=2em plus 12em \rightskip=\leftskip
  \parindent=0pt \parfillskip=0pt \spaceskip=.3333em \xspaceskip=.5em
  \pretolerance=9999 \tolerance=9999
  \hyphenpenalty=9999 \exhyphenpenalty=9999 }
\parskip=\medskipamount
\twelvepoint            
\overfullrule=0pt       
\def\preprintno#1{
 \rightline{\rm #1}}    
\def\author                     
  {\vskip 3pt plus 0.2fill \beginlinemode
   \singlespace \raggedcenter \twelvesc}
\def\affil                      
  {\vskip 3pt plus 0.1fill \beginlinemode
   \oneandahalfspace \raggedcenter \sl}
\def\abstract                   
  {\vskip 3pt plus 0.3fill \beginparmode
   \doublespace \narrower \noindent ABSTRACT: }
\def\endtitlepage               
  {\endpage                     
   \body}
\def\body                       
  {\beginparmode}               

\def\subhead#1{                 
  \vskip 0.1truein             
  {\raggedcenter #1 \par}
   \nobreak\vskip 0.1truein\nobreak}
\def\refto#1{$|{#1}$}           
\def\references                 
  {\subhead{References}         
   \beginparmode
   \frenchspacing \parindent=0pt \leftskip=1truecm
   \parskip=8pt plus 3pt \everypar{\hangindent=\parindent}}
\gdef\refis#1{\indent\hbox to 0pt{\hss#1.~}}    
\gdef\journal#1, #2, #3, 1#4#5#6{               
    {\sl #1~}{\bf #2}, #3, (1#4#5#6)}           
\def\refstylenp{                
  \gdef\refto##1{ [##1]}                                
  \gdef\refis##1{\indent\hbox to 0pt{\hss##1)~}}        
  \gdef\journal##1, ##2, ##3, ##4 {                     
     {\sl ##1~}{\bf ##2~}(##3) ##4 }}
\def\refstyleprnp{              
  \gdef\refto##1{ [##1]}                                
  \gdef\refis##1{\indent\hbox to 0pt{\hss##1)~}}        
  \gdef\journal##1, ##2, ##3, 1##4##5##6{               
    {\sl ##1~}{\bf ##2~}(1##4##5##6) ##3}}

\def\endreferences{\body}
\def\endpage                    
  {\vfill\eject}
\def\endpaper                   
  {\endmode\vfill\supereject}

\def\ref#1{Ref. #1}                     
\def\Ref#1{Ref. #1}                     

\def\m@th{\mathsurround=0pt }
\font\twelvesc=cmcsc10 scaled 1200
\def\cite#1{{#1}}
\def\(#1){(\call{#1})}
\def\call#1{{#1}}
\def\taghead#1{}
\def\leaderfill{\leaders\hbox to 1em{\hss.\hss}\hfill}
\def\twiddle{\lower.9ex\rlap{$\kern-.1em\scriptstyle\sim$}}
\def\bigtwiddle{\lower1.ex\rlap{$\sim$}}
\def\gtwid{\mathrel{\raise.3ex\hbox{$>$\kern-.75em\lower1ex\hbox{$\sim$}}}}
\def\ltwid{\mathrel{\raise.3ex\hbox{$<$\kern-.75em\lower1ex\hbox{$\sim$}}}}
\def\square{\kern1pt\vbox{\hrule height 1.2pt\hbox{\vrule width 1.2pt\hskip 3pt
   \vbox{\vskip 6pt}\hskip 3pt\vrule width 0.6pt}\hrule height 0.6pt}\kern1pt}
\catcode`@=11
\newcount\tagnumber\tagnumber=0

\immediate\newwrite\eqnfile
\newif\if@qnfile\@qnfilefalse
\def\write@qn#1{}
\def\writenew@qn#1{}
\def\w@rnwrite#1{\write@qn{#1}\message{#1}}
\def\@rrwrite#1{\write@qn{#1}\errmessage{#1}}

\def\taghead#1{\gdef\t@ghead{#1}\global\tagnumber=0}
\def\t@ghead{}

\expandafter\def\csname @qnnum-3\endcsname
  {{\t@ghead\advance\tagnumber by -3\relax\number\tagnumber}}
\expandafter\def\csname @qnnum-2\endcsname
  {{\t@ghead\advance\tagnumber by -2\relax\number\tagnumber}}
\expandafter\def\csname @qnnum-1\endcsname
  {{\t@ghead\advance\tagnumber by -1\relax\number\tagnumber}}
\expandafter\def\csname @qnnum0\endcsname
  {\t@ghead\number\tagnumber}
\expandafter\def\csname @qnnum+1\endcsname
  {{\t@ghead\advance\tagnumber by 1\relax\number\tagnumber}}
\expandafter\def\csname @qnnum+2\endcsname
  {{\t@ghead\advance\tagnumber by 2\relax\number\tagnumber}}
\expandafter\def\csname @qnnum+3\endcsname
  {{\t@ghead\advance\tagnumber by 3\relax\number\tagnumber}}

\def\equationfile{%
  \@qnfiletrue\immediate\openout\eqnfile=\jobname.eqn%
  \def\write@qn##1{\if@qnfile\immediate\write\eqnfile{##1}\fi}
  \def\writenew@qn##1{\if@qnfile\immediate\write\eqnfile
    {\noexpand\tag{##1} = (\t@ghead\number\tagnumber)}\fi}
}

\def\callall#1{\xdef#1##1{#1{\noexpand\call{##1}}}}
\def\call#1{\each@rg\callr@nge{#1}}

\def\each@rg#1#2{{\let\thecsname=#1\expandafter\first@rg#2,\end,}}
\def\first@rg#1,{\thecsname{#1}\apply@rg}
\def\apply@rg#1,{\ifx\end#1\let\next=\relax%
\else,\thecsname{#1}\let\next=\apply@rg\fi\next}

\def\callr@nge#1{\calldor@nge#1-\end-}
\def\callr@ngeat#1\end-{#1}
\def\calldor@nge#1-#2-{\ifx\end#2\@qneatspace#1 %
  \else\calll@@p{#1}{#2}\callr@ngeat\fi}
\def\calll@@p#1#2{\ifnum#1>#2{\@rrwrite{Equation range #1-#2\space is bad.}
\errhelp{If you call a series of equations by the notation M-N, then M and
N must be integers, and N must be greater than or equal to M.}}\else%
 {\count0=#1\count1=#2\advance\count1
by1\relax\expandafter\@qncall\the\count0,%
  \loop\advance\count0 by1\relax%
    \ifnum\count0<\count1,\expandafter\@qncall\the\count0,%
  \repeat}\fi}

\def\@qneatspace#1#2 {\@qncall#1#2,}
\def\@qncall#1,{\ifunc@lled{#1}{\def\next{#1}\ifx\next\empty\else
  \w@rnwrite{Equation number \noexpand\(>>#1<<) has not been defined yet.}
  >>#1<<\fi}\else\csname @qnnum#1\endcsname\fi}

\let\eqnono=\eqno
\def\eqno(#1){\tag#1}
\def\tag#1$${\eqnono(\displayt@g#1 )$$}

\def\aligntag#1\endaligntag
  $${\gdef\tag##1\\{&(##1 )\cr}\eqalignno{#1\\}$$
  \gdef\tag##1$${\eqnono(\displayt@g##1 )$$}}

\def\eqalignno#1{\displ@y \tabskip\centering
  \halign to\displaywidth{\hfil$\displaystyle{##}$\tabskip\z@skip
    &$\displaystyle{{}##}$\hfil\tabskip\centering
    &\llap{$\displayt@gpar##$}\tabskip\z@skip\crcr
    #1\crcr}}

\def\displayt@gpar(#1){(\displayt@g#1 )}

\def\displayt@g#1 {\rm\ifunc@lled{#1}\global\advance\tagnumber by1
        {\def\next{#1}\ifx\next\empty\else\expandafter
        \xdef\csname @qnnum#1\endcsname{\t@ghead\number\tagnumber}\fi}%
  \writenew@qn{#1}\t@ghead\number\tagnumber\else
        {\edef\next{\t@ghead\number\tagnumber}%
        \expandafter\ifx\csname @qnnum#1\endcsname\next\else
        \w@rnwrite{Equation \noexpand\tag{#1} is a duplicate number.}\fi}%
  \csname @qnnum#1\endcsname\fi}

\def\ifunc@lled#1{\expandafter\ifx\csname @qnnum#1\endcsname\relax}

\let\@qnend=\end\gdef\end{\if@qnfile
\immediate\write16{Equation numbers written on []\jobname.EQN.}\fi\@qnend}

\catcode`@=12
\refstyleprnp
\catcode`@=11
\newcount\r@fcount \r@fcount=0
\def\refreset{\global\r@fcount=0}
\newcount\r@fcurr
\immediate\newwrite\reffile
\newif\ifr@ffile\r@ffilefalse
\def\w@rnwrite#1{\ifr@ffile\immediate\write\reffile{#1}\fi\message{#1}}

\def\writer@f#1>>{}
\def\referencefile{
  \r@ffiletrue\immediate\openout\reffile=\jobname.ref%
  \def\writer@f##1>>{\ifr@ffile\immediate\write\reffile%
    {\noexpand\refis{##1} = \csname r@fnum##1\endcsname = %
     \expandafter\expandafter\expandafter\strip@t\expandafter%
     \meaning\csname r@ftext\csname r@fnum##1\endcsname\endcsname}\fi}%
  \def\strip@t##1>>{}}

\def\citeall#1{\xdef#1##1{#1{\noexpand\cite{##1}}}}
\def\cite#1{\each@rg\citer@nge{#1}}	

\def\each@rg#1#2{{\let\thecsname=#1\expandafter\first@rg#2,\end,}}
\def\first@rg#1,{\thecsname{#1}\apply@rg}	
\def\apply@rg#1,{\ifx\end#1\let\next=\relax
\else,\thecsname{#1}\let\next=\apply@rg\fi\next}

\def\citer@nge#1{\citedor@nge#1-\end-}	
\def\citer@ngeat#1\end-{#1}
\def\citedor@nge#1-#2-{\ifx\end#2\r@featspace#1 
  \else\citel@@p{#1}{#2}\citer@ngeat\fi}	
\def\citel@@p#1#2{\ifnum#1>#2{\errmessage{Reference range #1-#2\space is bad.}%
    \errhelp{If you cite a series of references by the notation M-N, then M and
    N must be integers, and N must be greater than or equal to M.}}\else%
 {\count0=#1\count1=#2\advance\count1
by1\relax\expandafter\r@fcite\the\count0,%
  \loop\advance\count0 by1\relax
    \ifnum\count0<\count1,\expandafter\r@fcite\the\count0,%
  \repeat}\fi}

\def\r@featspace#1#2 {\r@fcite#1#2,}	
\def\r@fcite#1,{\ifuncit@d{#1}
    \newr@f{#1}%
    \expandafter\gdef\csname r@ftext\number\r@fcount\endcsname%
                     {\message{Reference #1 to be supplied.}%
                      \writer@f#1>>#1 to be supplied.\par}%
 \fi%
 \csname r@fnum#1\endcsname}
\def\ifuncit@d#1{\expandafter\ifx\csname r@fnum#1\endcsname\relax}%
\def\newr@f#1{\global\advance\r@fcount by1%
    \expandafter\xdef\csname r@fnum#1\endcsname{\number\r@fcount}}

\let\r@fis=\refis			
\def\refis#1#2#3\par{\ifuncit@d{#1}
   \newr@f{#1}%
   \w@rnwrite{Reference #1=\number\r@fcount\space is not cited up to now.}\fi%
  \expandafter\gdef\csname r@ftext\csname r@fnum#1\endcsname\endcsname%
  {\writer@f#1>>#2#3\par}}

\def\ignoreuncited{
   \def\refis##1##2##3\par{\ifuncit@d{##1}%
     \else\expandafter\gdef\csname r@ftext\csname
r@fnum##1\endcsname\endcsname%
     {\writer@f##1>>##2##3\par}\fi}}

\def\r@ferr{\endreferences\errmessage{I was expecting to see
\noexpand\endreferences before now;  I have inserted it here.}}
\let\r@ferences=\references
\def\references{\r@ferences\def\endmode{\r@ferr\par\endgroup}}

\let\endr@ferences=\endreferences
\def\endreferences{\r@fcurr=0
  {\loop\ifnum\r@fcurr<\r@fcount
    \advance\r@fcurr by 1\relax\expandafter\r@fis\expandafter{\number\r@fcurr}%
    \csname r@ftext\number\r@fcurr\endcsname%
  \repeat}\gdef\r@ferr{}\global\r@fcount=0\endr@ferences}

\let\r@fend=\endpaper\gdef\endpaper{\ifr@ffile
\immediate\write16{Cross References written on []\jobname.REF.}\fi\r@fend}

\catcode`@=12

\citeall\refto		
\citeall\ref		%
\citeall\Ref		%

\referencefile

\font\titlefont=cmr10 scaled\magstep3
\def\bigtitle                      
  {\null\vskip 3pt plus 0.2fill
   \beginlinemode \doublespace \raggedcenter \titlefont}

\

\preprintno{UFIFT-HEP-94-20}
\centerline{INTRODUCTORY LECTURES ON LOW ENERGY SUPERSYMMETRY}
\vskip .5cm
\centerline{P. Ramond}
\centerline{Institute for Fundamental Theory, Department of
Physics}
\centerline{University of Florida, Gainesville, FL 32611}
\vskip 2cm
In these lectures, I cover the most elementary aspects of $N=1$
supersymmetry, and its application to low energy phenomenology. Since
there is no end to the subject, I decided not to cover supergravity,
rather concentrate on the basic techniques of global supersymmetry, in
the context of the $N=1$ Standard Model. I do discuss, but only
cursorily, the all important question of supersymmetry breaking. The
lectures are organized as follows:
\vskip .5cm
I-) Motivation
\vskip .3cm
II-) Tools: The Chiral and Vector Supermultiplets
\vskip .3cm
III-) The Minimal $ N=1$ Standard Model
\vskip .3cm
IV-) More Tools: Supersymmetry Breaking
\vskip 2cm
\centerline{\bf I-) MOTIVATION}
\vskip .3cm
The $N=0$ Standard Model is a very compact model, described by three
gauge groups, and nineteen parameters. The quantum numbers of the
fermions strongly suggest that the three gauge groups are part of a more
integrated structure, but its parameters range all over over the place,
and show few if any discernable patterns.

We also know that the model becomes inconsistent at distances shorter
than the Planck length because of the divergent nature of quantum
gravity. Thus we should regard the model as an effective theory, valid
at larger distances, and view the Planck scale as  Nature's own
ultraviolet cutoff.

It is natural to ask if the standard model can be described in simpler
terms at shorter distances. The only tool at our disposal is the
renormalization group. If we assume no new physics at shorter distances, the
renormalization group allows us to extrapolate the parameters of the
standard model to the deep ultraviolet, and look for the patterns
suggested by the quantum numbers.

To start, the inverse of the fine structure constants for the three
gauge groups evolve linearly with the logarithm of the scale
$${d\alpha^{-1}_i\over dt}={1\over 2\pi}b_i\ ,$$
where  $b_i=(-{41\over
10},{19\over 6},7)$ for $U(1)$, $SU(2)$ and $SU(3)$, respectively,
and
$t=\ln(\mu/\mu_0)$, $\mu$ being the scale normalized to an
arbitrary
scale $\mu_0$.

Using low energy data as boundary conditions, the two weak fine
structure constants meet at a scale of $10^{13}$ GeV, with a value
$\alpha^{-1}=43$. At the same scale, the QCD fine structure constant
weighs in at $\alpha^{-1}_3=38$. Thus the road to Grand Unified Theories
(GUTs), the apparent unification suggested by the quantum numbers, is
not achieved in the $N=0$ Standard Model.

Only the hypercharge couplings becomes non-perturbative at shorter
distances, but only well beyond the Planck scale, as does the electric
coupling in QED.

The Yukawa couplings also behave supinely in the ultraviolet.
Generically their renormalization group equations are ruled by two
competing effects. One is the Yukawa couplings themselves, which tend to
make the coupling blow up at short distances, the other from the gauge
couplings, does the opposite. In the $N=0$ Standard Model, the QCD gauge
couplings dominate, and the Yukawa couplings gently settle to
non-perturbative values. To give an example, the top Yukawa coupling
varies according to

$${dy_t\over dt}={y_t^{}\over
16\pi^2}({9\over 2}y_t^2-8g_3^2)\ ,$$
which is negative around $M_Z$. Although both the Yukawa and gauge
coupling decrease, the $\beta$ function does not change sign at shorter
distances. As for the leptons, their Yukawas are too small to dominate
the electroweak couplings.

This leaves us with the Higgs self-coupling. Its $\beta$ function is
ruled by two effects which work in opposite directions. The contribution
from the self coupling itself forces it to blow up, while the fermion
loop correction works in the opposite direction; in addition there are
contributions from the gauge couplings, but they are small at
experimental energies, since the Higgs has no color. Neglecting the
gauge contributions, we have

$${d\lambda\over dt}={1\over
16\pi^2}(12\lambda^2+12y_t^2\lambda-12y_t^4)\ .$$
It is convenient to discuss the behavior of this coupling in terms of
the Higgs mass, which is proportional to $\sqrt\lambda$. If the Higgs
mass is large, then so is $\lambda$, and the first term dominates. It
quickly drives the coupling to non-perturbative values. The larger the
Higgs mass, the sooner it blows up; since there is no evidence of
non-perturbative behavior, this sets an upper bound on the Higgs mass.
This is called the triviality bound. If the Higgs mass is found to be
above $200$ GeV, it is certain that non-perturbative physics is present
in the TeV region. Since there is no evidence of strong coupling in the
electroweak model at experimental scales, the Higgs cannot be
arbitrarily massive; it must be lighter than 600 to 800 GeV.

On the other hand, if the Higgs mass is small, the last term which does
not depend on $\lambda$ takes over and drives $\lambda$ towards negative
values. However a negative $\lambda$ means that the potential is no
longer stable, since its potential becomes unbounded below. Larger field
configurations are favored, which takes us beyond perturbation theory.
With the recent value of the top quark mass, if the Higgs is less than
$150$ GeV, the standard model description leaves perturbation theory at
some scale below the Planck scale. For a Higgs mass around $100$ GeV,
this $``$instability bound" sets in around one TeV. It can be cured in a
variety of ways, say by adding new degrees of freedom. Supersymmetry is
one theory which obviates this bound by eliminating the $\lambda$
coupling altogether!

We conclude that in the $N=0$ standard model, non-perturbative
physics below the Planck scale is expected for a wide range of the Higgs
mass. Should there be non-perturbative physics at some scale below the
Planck mass, we must view the $N=0$ standard model as an effective
theory with that scale as a cut-off, rather than the Planck scale.

The dependence of the standard model parameters on the cut-off is
illuminating. The fermion masses, for instance, depend on the logarithm
of the cut-off, as can be seen by evaluating the one-loop correction to
the mass. The reason for this mild dependence is chiral
symmetry; it softens the degree of divergence of the diagram. This
protective symmetry works because the theory becomes chirally invariant
in the limit of massless fermions.

Another nearly massless (on the scale of Planck mass) particle is the
Higgs scalar, but its dependence on the cut-off is linear! There is no
symmetry to protect it. It is therefore natural to expect that the mass
of the Higss is of the order of the cut-off of the $N=0$ standard model,
times some coupling constant. Thus the expectation that the natural
cut-off is of the order of TeVs. Generically, theories which detail this
possibility are called {\it technicolor} theories. In such theories, the
longitudinal W bosons interact strongly under the strong technicolor
force, in direct analogy to the pions in the strong interactions.
We note with some amusement, that historically,  attempts to formulate
such theories have led to string theories, which then led to superstring
theories ({\it Plus \c ca change,...}).

One could conceivably avoid this conclusion if the parameters of the
theory are so finely tuned that the cut-off dependence of the Higgs mass
is relegated to higher loop effects. Since we do not know the origin of
the parameters, this is a logical, albeit unfair possibility to keep in
mind. It could be that the parameters are at a fixed point of a
non-linear mother theory, with fractal-like relations among themselves.
I thought I would mention this to open your ears to this possible
application of chaotic phenomena.

One can demonstrably avoid strong coupling at low energies, by generalizing
the $N=0$
standard model to supersymmetry, which we call the $N=1$ standard model.
Supersymmetry links the Higgs field to a chiral fermion of equal mass.
Then the chiral symmetry which protects the fermion protects the Higgs
as well, and results in the same logarithmic dependence of the Higgs
mass on the cut-off. As long as the boson-fermion supersymmetry is
unbroken, that is. The lack of evidence of such symmetry in the low
energy world, indicates that it must be broken, and the trick is to
break it at a scale that is not too high tlest it unravels
its salutory effect.
All that supersymmetry does is to allow all the couplings to remain
perturbative all the way to at near the Planck mass. Then, the $N=1$
standard model is predictive to the Planck scale, but none of the
mysteries associated with the breaking of the electroweak symmetry have
been explained; they have just been shuffled in the yet to come
explanation of the breaking of supersymmetry. However, in generic
situations, it turns out that supersymmetry breaking induces electroweak
breaking. All that remains is to explain how supersymmetry breaking
comes about, and at what scale.

\vskip 2cm
\centerline{\bf II-) SUPERSYMMETRY TOOLBOX}
\vskip .5cm

\noindent This is the first of a series of sections about $N=1$
supersymmetry. It is not meant to be complete, but rather helpful in
presenting the relevant facts. In the absence of gravity, supersymmetry
employs two collections of fields, arranged in supersymmetric
multiplets. The first, called the chiral or Wess-Zumino supermultiplet,
consists of one left-handed Weyl spinor and a complex scalar field, and
it serves as a generalization of the fermion and Higgs fields of the
Standard Model. The second, called the gauge supermultiplet, contains
the gauge vector bosons, as well as their supersymmetric partners, the
gauginos. The $N=1$ Standard Model is described by these supermultiplets
in interaction with one another.
\vskip 1cm
\centerline{The Chiral Supermultiplet}
\vskip .5cm
\noindent We start with a brief description of our notation and
conventions. The student unfamiliar with these is encouraged to consult
standard texts on Advanced Quantum Mechanics and Field Theory.

In four space-time dimensions, the algebra of the Lorentz group is
isomorphic to that of $SU_2 \times SU_2$ (up to factors of $i$), the
first generated by $\vec J + i \vec K$, the second by $\vec J - i \vec
K; \vec J$ are the generators of angular momentum and $\vec K$ are the
boosts.  Thus these two $SU_2$ are seen to be connected by complex
conjugation $(i \rightarrow -i)$ and/or parity $(\vec K \rightarrow
-\vec K, \vec J \rightarrow \vec J)$, and are therefore left invariant
by the combined operation of CP.  In accordance with this algebraic
structure, spinor fields appear in two varieties, left-handed spinors,
transforming under the first $SU_2$ as spin ${1\over 2}$
representations, and right-handed spinors transforming only under the
second $SU_2$.  They are represented by two-component complex spinor
fields, called Weyl spinors,

$$\psi^{}_L(x) \sim ({\bf 2},{\bf 1})\ ,\qquad
\psi^{}_R(x) \sim ({\bf 1},{\bf 2}) \ .$$
The spinor fields must be taken to be anticommuting Grassmann variables,
in accordance with the Pauli exclusion principle. Their Lorentz
transformation properties can be written in terms of the Pauli spin
matrices

$$\sigma_1 = \left(\matrix{0&1\cr  1&0\cr }\right)\ ,\ \ \sigma_2 =
\left(\matrix{0&-i\cr  i&0\cr }\right)\ ,\ \ \sigma_3 =
\left(\matrix{1&0\cr  0&-1\cr }\right)\ ,$$
which satisfy
$$\sigma_i \sigma_j = \delta_{ij} + i\epsilon_{ijk} \sigma_\kappa\ .$$

The action of the Lorentz group on the spinor fields is
$$\psi^{}_{L,R} \rightarrow
{\bf\Lambda}^{}_{L,R}\psi^{}_{L,R}\equiv
e^{{i\over 2} \vec \sigma \cdot(\vec
\omega \mp i \vec \nu)} \psi^{}_{L,R}\ ,$$
where $\vec \omega$ and $\vec \nu$ are the real rotation and boost
angles, respectively. This corresponds to the representation where
$${\vec J}={{\vec\sigma}\over 2}\ ;\qquad {\vec
K}=-i{{\vec\sigma}\over
2}\ .$$
It is possible to make left-handed spinors out of
right-handed antispinors, and vice-versa.  One checks that
$$\eqalign{\overline \psi^{}_L &\equiv \sigma_2 \psi_R^\ast \sim
({\bf
2},{\bf 1})\ ,\cr
\overline \psi_R &\equiv \sigma_2 \psi_L^\ast \sim ({\bf 1},{\bf
2})
\  .\cr }$$
Under charge conjugation the fields behave as
$$C\ : \ \psi^{}_L \rightarrow \sigma_2 \psi_R^\ast\ ,~~~~
\psi^{}_R \rightarrow - \sigma_2 \psi_L^\ast\ .$$
Under parity
$$P\ : \ \psi^{}_L \rightarrow \psi^{}_R\ ,~~~~
\psi^{}_R \rightarrow \psi^{}_L\ .$$

This purely left-handed notation is convenient to describe fermions that
interact by the weak interactions which violate parity. For instance the
neutrinos appear only as left-handed fields, while the antineutrinos are
purely right-handed. On the other hand, fermions which interact in a
parity invariant way as in QED and QCD, have both left- and right-handed
parts. In that case, it is far more convenient to use the Dirac
four-component notation. The fields $\psi^{}_L$ and $\psi^{}_R$ are put
together into a four-component Dirac spinor (in the Weyl representation)

$$\Psi = \left( \matrix{\psi^{}_L\cr  \psi^{}_R\cr }\right)\ ,$$
on which the operation of parity is well-defined.  In this
representation, called the Weyl representation, the anticommuting Dirac
matrices are (in $2\times 2$ block form)

$$\gamma_0 = \left(\matrix{0&1\cr  1&0\cr }\right)\ ,\  \gamma_i=
\left(\matrix{0&-\sigma_i\cr  \sigma_i &0\cr }\right)\ ,$$
$$\gamma_5 = i\gamma^0 \gamma^1 \gamma^2 \gamma^3 =
\left(\matrix{1&0\cr  0&-1\cr }\right)\ .$$
Since one can generate right-handed fields starting from left-handed
ones, it suffices to consider only polynomials made out of left-handed
fields.

We will be doing many manipulations in Weyl language, and it is useful
to see how the Fierz transformations read. Let $\zeta$ and $\eta$ be
complex two-component Weyl spinors, each transforming as ({\bf 2},{\bf
1}) of the Lorentz group.  Thus
the convenient Fierz decompositions
$$\zeta \eta^T\sigma_2 = - {1\over 2} \sigma_2 \eta^T\sigma_2
\zeta -
{1\over 2} \sigma^i \eta^T\sigma_2
\sigma^i \zeta\ ,$$
corresponding to
$$({\bf 2},{\bf 1}) \otimes ({\bf 2},{\bf 1}) = ({\bf 1},{\bf 1})
\oplus ({\bf 3},{\bf 1})\ .$$
As the combinations $\sigma_2
\zeta^\ast$ and $\sigma_2\eta^\ast$ transform according to the ({\bf
1},{\bf 2}) representation, we also have

$$\zeta \eta^\dagger = - {1\over 2} \eta^\dagger \zeta - {1\over 2}
\sigma^i \eta^\dagger \sigma^i \zeta\ ,$$
corresponding to
$$({\bf 2},{\bf 1}) \otimes ({\bf 1},{\bf 2}) = ({\bf 2},{\bf 2})\
.$$
The right hand side of this equation does indeed correspond to the
vector representation, as we can see by introducing the matrices
$$\sigma^\mu = (\sigma^0=1, \sigma^i)\ ;~~~~~\overline \sigma^\mu
=(\sigma^0=1,-\sigma^i) \ ,$$
in terms of which we rewrite
$$\zeta \eta^\dagger = - {1\over 2} \overline \sigma^\mu
\eta^\dagger
\sigma_\mu \zeta\ .$$
For the Pauli matrices, we have
$$\sigma_2 \sigma^i \sigma_2 = -\sigma^{iT} = -\sigma^{i\ast}\ ,$$
so that
$$\sigma_2 \overline \sigma^\mu \sigma_2 = \sigma^{\mu T}\ .$$

The simplest set of fields on which $N=1$ supersymmetry is realized is
the chiral or Wess-Zumino multiplet which contains the three fields,

$$\eqalign{\varphi (x) &\ ,~{\rm a~complex~scalar}\ ,\cr\noalign{\vskip 0.2cm}
\psi (x) &\ ,~ {\rm a~Weyl~spinor}\ ,\cr\noalign{\vskip 0.2cm}
F(x) &,\ ~{\rm a~complex~auxiliary~field}\ .\cr\noalign{\vskip
0.2cm}}$$
The Lagrangian density is given by

$${\cal L}_0^{WS} \equiv \partial_\mu \varphi^\ast \partial^\mu
\varphi + \psi^\dagger \sigma^\mu \partial_\mu \psi + F^\ast F\ ;$$
it is invariant, up to a surface term, under the following
transformations

$$\eqalign{\delta \varphi &=  \alpha^T\sigma_2
 \psi\
,\cr\noalign{\vskip 0.2cm}
\delta \psi &= \alpha F - \overline \sigma^\mu
\sigma_2\alpha^*\partial_\mu \varphi\ ,\cr
\delta F &= -  \alpha^\dagger \sigma^\mu \partial_\mu
\psi\ .\cr\noalign{\vskip 0.2cm}}$$
Here $\alpha$ is the parameter of the supersymmetry transformation; it
is a Weyl spinor.  Note that we treat Grassmann variables as plain
anticommuting numbers, so that for any two of them

$$(\zeta^T\sigma_2
\chi)^\ast = \zeta^\dagger \sigma_2^\ast
\chi^\ast = - \zeta^\dagger\sigma_2\chi^\ast\ ;$$
this is the reason there is no $i$ in front of the fermion kinetic
term.  Take note of the following: $\varphi$ and $\psi$ have the canonical
dimensions, $-1$ and $-3/2$, but F has the non-canonical dimension
of $-2$, and $\alpha$ has dimension $1/2$.  Also, F transforms as
a total divergence.

Under two supersymmetry transformations, labelled $\delta_1$ and
$\delta_2$, with parameters $\alpha_1$ and $\alpha_2$, we find
that
$$[\delta_1, \delta_2] \ast = (\alpha_1^\dagger \sigma^\mu
\alpha^{}_2 - \alpha_2^\dagger \sigma^\mu \alpha^{}_1) \partial_\mu
\ast \ ,$$
where $\ast$ stands for any of the three fields, $\varphi,~\psi,$
and F.  This equation shows that the result of two supersymmetry
transformations is just a translation by the amount
$$\delta x^\mu = (\alpha_1^\dagger \sigma^\mu \alpha^{}_2 -
\alpha_2^\dagger \sigma^\mu \alpha^{}_1)\ ,$$
recalling that $P_\rho = -i\partial_\rho$ is the generator of
translations. Thus supersymmetry transformations are the square root of
translations, and we will see later how the Poincar\' e group is altered
to accomodate these new transformations.

Let us verify this equation for one of the fields. For example

$$\eqalign{\delta_1 \delta_2 F &= - \alpha_2^\dagger
\sigma^\mu \partial_\mu \delta_1 \psi\ ,\cr\noalign{\vskip 0.2cm}
&= -\alpha_2^\dagger
\sigma^\mu \alpha_1 \partial_\mu F - \alpha_2^\dagger \sigma^\mu
\overline \sigma^\rho\sigma_2 \alpha_1^*\partial_\mu
\partial_\rho \varphi\ .\cr\noalign{\vskip 0.2cm}}$$
Now because of the symmetry of $\partial_\mu \partial_\rho
\varphi$, we can set
$$\sigma^\mu \overline \sigma^\rho = {1\over 2}
(\sigma^\mu \overline
\sigma^\rho + \sigma^\rho \overline \sigma^\mu)= g^{\mu\rho}\ ,$$
leading to
$$\delta_1 \delta_2 F = - \alpha_2^\dagger \sigma^\mu \alpha^{}_1
\partial_\mu F - \alpha_2^\dagger\sigma_2\alpha_1^*
g^{\mu\nu} \partial_\mu \partial_\nu \varphi\ .$$
Now $\alpha_2^\dagger\sigma_2\alpha_1^*$ is symmetric under
the $(1 \leftrightarrow 2)$ interchange, and drops out from the
commutator, giving the desired result
$$[\delta_1, \delta_2] F = (\alpha_1^\dagger \sigma^\mu
\alpha^{}_2
- \alpha_2^\dagger \sigma^\mu \alpha^{}_1) \partial_\mu F\ .$$
The other two expressions for $\varphi$ and $\psi$ work out in a
similar way, making use of Fierz identities when applied to $\psi$.

All these results can be neatly summarized by introducing a
two-component Weyl
Grassmann variable $\theta$.  We introduce the superfield
$\Phi(x,\theta)$ which depends only on $\theta$, its most general
expansion is
$$\Phi(x,\theta) = \varphi (x) + \theta^T\sigma_2
\psi(x) +
{1\over 2}\theta^T\sigma_2
\theta F(x)\ ,$$
such that its change under supersymmetry can be obtained by acting on
the fields,
$$\delta \Phi = \delta \varphi + \theta^T\sigma_2
\delta \psi +
{1\over 2}\theta^T\sigma_2
\theta \delta F\ ,$$
or as operator acting on the coordinates
$$\delta \Phi = \left [ \alpha^T\sigma_2
{\partial \over
\partial \theta} + \alpha^\dagger \sigma^\mu \theta
\partial_\mu \right ] \Phi\ ,$$
where we have introduced the Grassmann derivative, defined through
$${\partial\over \partial \theta} \theta^T\sigma_2
= 1\ .$$
Note that expressing the supersymmetry transformations as generated by
derivative operators enables us to derive the commutator formula in a
much more elegant way. It also enables us to see why the change in the
coefficient of $\theta^T\sigma_2
\theta$ is a total divergence: it can only
come from the term linear in $\theta$ in the generator, which contains
the space-time derivative, acting on the term linear in $\theta$ in the
superfield.

We can express the effect of a supersymmetry transformation on the
chiral superfield in another way, namely
$$\Phi (x^\mu, \theta) \rightarrow \Phi (x^\mu + \alpha^\dagger
\sigma^\mu \theta, \theta + \alpha)\ ;\qquad{\partial \Phi\over
\partial
\theta^\ast} = 0\ .$$
While formally pleasing, we note that the change in $x^\mu$ is
not real.  If we decompose it into its real plus imaginary parts,
$$\alpha^\dagger
\sigma^\mu \theta={1\over 2}(\alpha^\dagger \sigma^\mu \theta -
 \theta^\dagger
\sigma^\mu \alpha)+{1\over 2}(\alpha^\dagger \sigma^\mu \theta +
 \theta^\dagger\sigma^\mu \alpha)\ ,$$
we note that the imaginary part can itself be written as the change of half
the quantity
$\theta^\dagger\sigma_\mu\theta$
under a shift of the Grassmann variables.

Such considerations lead us to construct the superfield

$$V(x^\mu, \theta, \theta^\ast) = \Phi^\ast (x^\mu +
{1\over 2}\theta^\dagger \sigma^\mu \theta, \theta) \Phi(x^\mu +
{1\over 2}\theta^\dagger \sigma^\mu \theta, \theta)\ .$$
It is manifestly real

$$V^\ast (x^\mu, \theta, \theta^\ast) = V(x^\mu, \theta,
\theta^\ast)\ \ ,$$
and transforms under supersymmetry in an aesthetic way, namely
the change in the coordinate $x_\mu$ is real:

$$V(x^\mu, \theta, \theta^\ast) \rightarrow V(x^\mu +
{1\over 2}(\alpha^\dagger \sigma^\mu \theta - \theta^\dagger \sigma^\mu
\alpha), \theta + \alpha, \theta^\ast + \alpha^\ast)\ .$$
However this superfield depends on the Grassmann variables and their
conjugates, but the change in the coordinate is now real. To verify this
equation, we start from the chiral superfield

$$\eqalign{\Phi (x^\mu + {1\over 2}\theta^\dagger \sigma^\mu \theta,
\theta) = &\varphi(x) + \theta^T\sigma_2
\psi(x) +
{1\over 2}\theta^T\sigma_2
\theta F(x)\cr\noalign{\vskip 0.2cm}
&+{1\over 2}\theta^\dagger\sigma^\mu\theta\partial_\mu\varphi(x)-{1\over 4}
\theta^T\sigma_2
\theta \theta^\dagger
\sigma^\mu \partial_\mu \psi (x)
+ {1\over 16}
\vert\theta^T\sigma_2
\theta\vert^2 \partial^\mu
\partial_\mu \varphi(x)\ ,\cr\noalign{\vskip 0.2cm}}$$
where we have used some Fierzing and the identity

$$\theta^\dagger \sigma^\mu \theta \theta^\dagger \sigma^\nu\theta
= {1\over 2} g^{\mu\nu} \vert\theta^T\sigma_2
\theta\vert^2
\ .$$
It follows from the above that the real superfield is given by

$$\eqalign{V(x, \theta, \theta^\ast) = &\varphi^\ast(x) \varphi(x)
+ [\theta^T\sigma_2
 \psi \varphi^\ast - \theta^\dagger
\sigma_2 \psi^\ast \varphi]\cr\noalign{\vskip 0.2cm}
&+{1\over 2}[\theta^T\sigma_2
 \theta \varphi^\ast F - \theta^\dagger
\sigma_2
\theta^\ast \varphi F^\ast + \theta^\dagger \sigma^\mu \theta
(\varphi^\ast
\partial_\mu \varphi - \partial_\mu \varphi^\ast \varphi -
 \psi^\dagger
\sigma_\mu \psi)]\cr\noalign{\vskip 0.2cm}
&-{1\over 4}\theta^\dagger \sigma_2 \theta^\ast (2F^\ast \psi^T
\sigma_2 +  \varphi \partial_\mu \psi^\dagger \sigma^\mu
- \partial_\mu \varphi \psi^\dagger \sigma^\mu)
\theta\cr\noalign{\vskip 0.2cm}
&-{1\over 4}\theta^T\sigma_2
 \theta \theta^\dagger (2\sigma_2
\psi^\ast F +  \varphi^\ast \sigma^\mu \partial_\mu \psi
- \sigma^\mu \psi \partial_\mu
\varphi^\ast)\cr\noalign{\vskip 0.2cm}
+{1\over 8}&\vert\theta^T\sigma_2
 \theta\vert^2
(2F^\ast F -  \partial_\mu \varphi^\ast \partial^\mu
\varphi +
{1\over 2} (\varphi^\ast \partial^\mu \partial_\mu \varphi +
\varphi \partial^\mu
\partial_\mu \varphi^\ast)
+ \psi^\dagger \sigma^\mu \partial_\mu \psi -
\partial_\mu \psi^\dagger \sigma^\mu \psi)\ .\cr\noalign{\vskip
0.2cm}}$$
The alert student will recognize the last term as the Lagrange
density, plus an overall divergence.

You can verify the transformation law. We show it to hold on a subset of
terms of the form $\partial_\mu \varphi \varphi^\ast$. On the one hand,
we get

$$\eqalign{V(x^\mu + {1\over 2}(\alpha^\dagger \sigma^\mu \theta &-
\theta^\dagger
\sigma^\mu \alpha), \theta + \alpha, \theta^\ast +
\alpha^\ast)\cr\noalign{\vskip 0.2cm}
=&\cdots+{1\over 2}\varphi^\ast (\alpha^\dagger \sigma^\mu \theta -
\theta^\dagger\sigma^\mu \alpha) \partial_\mu \varphi +
{1\over 2}(\alpha^\dagger \sigma^\mu \theta + \theta^\dagger \sigma^\mu
\alpha) \varphi^\ast \partial_\mu \varphi +\cdots\cr\noalign{\vskip 0.2cm}
=& \alpha^\dagger \sigma^\mu \theta \varphi^\ast \partial_\mu
\varphi \ .\cr\noalign{\vskip 0.2cm}}$$
On the other hand, by varying the fields directly, we obtain the very
same term
$$\eqalign{\theta^T\sigma_2 \delta \psi \varphi^\ast
&= -\theta^T\sigma_2 \overline \sigma^\mu \sigma_2
\alpha^\ast
\partial_\mu \varphi \varphi^\ast + \ ,\cr\noalign{\vskip
0.2cm}
&=  \alpha^\dagger \sigma^\mu \theta \partial_\mu \varphi
\varphi^\ast \ .\cr\noalign{\vskip 0.2cm}}$$

We can also express the supersymmetric change on the real superfield as
resulting from the action of differential operators, namely

$$\eqalign{\delta V &= \{\alpha^T\sigma_2  {\partial\over
\partial \theta} - \alpha^\dagger \sigma_2 \left({\partial\over
\partial \theta}\right)^\ast + {1\over 2}(\alpha^\dagger \sigma^\mu \theta -
\theta^\dagger \sigma^\mu \alpha) \partial_\mu\} V(x, \theta,
\theta^\ast)\ ,\cr\noalign{\vskip 0.2cm}
&= \{\alpha^T\sigma_2
({\partial\over \partial \theta} +{1\over 2} \overline \sigma^\mu \sigma_2
\theta^\ast \partial_\mu) - \alpha^\dagger \sigma_2 (\left({\partial\over
\partial \theta}\right)^\ast -{1\over 2} \sigma_2 \sigma^\mu \theta
\partial_\mu)\}
V(x,\theta, \theta^\ast)\ ,\cr\noalign{\vskip 0.2cm}}$$
where we have used the identity
$$\theta^\dagger \sigma^\mu
\alpha=-\alpha^T\sigma_2
\overline\sigma^\mu\hat\theta^\dagger \ .$$
Introduce the generators of supersymmetry
$$\eqalign{Q &= {\partial\over \partial \theta} + {1\over 2}\overline
\sigma^\mu
\sigma_2 \theta^\ast \partial_\mu\ ,\cr\noalign{\vskip 0.2cm}
Q^\ast &= \left({\partial\over \partial \theta}\right)^\ast - {1\over 2}
\sigma_2 \sigma^\mu \theta \partial_\mu\ ,\cr\noalign{\vskip 0.2cm}}$$
to write the change in the real superfield
$$\delta V= (\alpha^T\sigma_2   Q
- \alpha^\dagger \sigma_2 Q^\ast) V(x, \theta, \theta^\ast)\ ,$$

The supersymmetry generators satisfy the anticommutation relations
$$\eqalign{\{Q, Q\} &= \{Q^\ast, Q^\ast\} = 0\ ,\cr\noalign{\vskip
0.2cm}
\{Q, Q^\ast \} &= \overline \sigma^\mu \partial_\mu\
.\cr\noalign{\vskip 0.2cm}}$$
When added to  the generators of the Poincar\' e group, these generators
form the super-Poincar\' e group, and the particles described by
supersymmetry must form irreducible representations of this supergroup.
We first note that the supersymmetry generators commute with
translations,
$$[ Q,P_\mu ]=0\ ,$$
hence with $P_\mu P^\mu$, the Casimir operator whose value is the mass
squared. Since the Poincare group is a subgroup, any representation of
the supergroup contains several representations of the Poincar\' e group of
the same mass.

It is simplest to start with massless representations.
The massless representations of the Poincar\' e group are labelled by
the helicity $\lambda$ which runs over positive and negative integer and
half-integer values. In local field theory, each helicity state
$\vert\lambda>$ is accompanied by its CPT conjugate $\vert-\lambda>$.
For example, the left polarized photon $\vert\lambda=+1>$ and its CPT
conjugate the right polarized photon $\vert\lambda=-1>$.

Let us go to the infinite momentum frame $P_0=P_3\ne 0$, where the
supersymmetry algebra reduces to the Clifford algebra
$$\{Q_1,Q^\ast_1\}=iP_0\ ,$$ all other anticommutators being zero. It
follows that we have just one supersymmetry operator and its conjugate,
acting like a raising operator. Thus starting with any state
$\vert~\lambda>$, we generate only one other state
$Q^\ast\vert~\lambda>$, which has helicity $~\lambda+1/2$. A repeated
application of the raising operator yield zero since $Q^2_1=0$. Hence
there are no other states. This yields the only irreducible
representation for supersymmetry of massless states: two states,
differing by half a unit of helicity. (I first learned this elegant proof
from Gell-Mann and Ne\'eman in 1975)

The Wess-Zumino multiplet corresponds to the representation
$\vert~0>\oplus\vert~1/2>$, together with its CPT conjugate
$\vert~ 0>\oplus\vert~ -1/2>$, which describes one Weyl fermion and two
scalar degrees of freedom.

In the same way we can expect the gauge multiplet which contains the
states  $\vert~ 1>\oplus\vert~ 1/2>$, together with their conjugates: a
vector particle and a Weyl fermion. There are other representations,
such as the graviton-gravitino combination, made up of $\vert~
2>\oplus\vert~ 3/2>$, plus conjugate. These are all realized in local
field theory. The number of bosonic and fermionic degrees of freedom
match exactly. For instance, the chiral multiplet has, using the
equations of motion, two fermionic degrees of freedom, exactly matched
by the complex scalar field. If the equations of motion are not used,
the number of fermions doubles, but the excess fermions is exactly
matched by adding two boson fields, the complex auxiliary field $F$.

Massive multiplets can always be obtained by assembling massless
multiplets, {\it \` a la} Higgs.

The real supermultiplet is highly reducible. It can be checked that
the covariant derivative operator

$$D \equiv {\partial\over \partial \theta} - {1\over 2}\overline \sigma^\mu
\sigma_2 \theta^\ast \partial_\mu\ ,$$
and its complex conjugate anticommute with the generators of
supersymmetry. By requiring that they vanish on the real superfield, we
obtain the chiral superfield. I leave this an an exercise to the
hardiest among you.

This notation in terms of Grassmann variables allows us to write
supersymmetric invariants in a very elegant way. We have already noted
that the highest component of a superfield transforms as a
four-divergence, so that its integral over space-time is supersymmetric
invariant.  We can define integration over the $\theta$ Grassmann
variables

$$\int d\theta=0\ ,~\int d\theta\theta=1\ ;$$
note that since $\theta$ has dimension 1/2, $d\theta$ has the opposite
dimension, -1/2. Integration enables us to rewrite the invariant in the
form

$$\int d^4 x \int d^2 \theta \Phi (x, \theta) = \int d^4 x F\ .$$
However, any product of $\Phi (x, \theta)$ is
itself a chiral superfield. To see this, note that by Fierzing,
$$\theta\theta^T\sigma_2\theta  = - {1\over 2}
\sigma_2 \theta^T\sigma_2 \theta \sigma_2 \theta = - {1\over 2}
\theta^T\sigma_2\theta \theta\ ,$$
so that
$$\theta^T\sigma_2 \theta \theta = 0\ ,$$
Thus the polynomial expansion in $\theta$, is exactly of the same form
as that of $\Phi$.

It follows that for any number of chiral superfields $\Phi_a,~ a=1,\dots
,N$, all the quantities

$$\int d^4x \int d^2 \theta \Phi_{a_i}\cdots
\Phi_{a_n}~~~{\rm for~all}~a_i~{\rm and}~ n\ ,$$
are supersymmetric invariants.  In terms of components, the lowest
polynomials are given by
$$\eqalign{m \int d^2 \theta \Phi_1 \Phi_2 &= m(\varphi_1 F_2 +
\varphi_2 F_1 - \psi_1^T \sigma_2 \psi_2)\ ,\cr
\lambda \int d^2 \theta \Phi_1 \Phi_2 \Phi_3 &= \lambda
(\varphi_1 \varphi_2 F_3 + \varphi_1 F_2 \varphi_3 + F_1
\varphi_2 \varphi_3\cr ~~~~~~~& - \varphi_1 \psi_2^T \sigma_2
\psi_3 -
\varphi_2 \psi_1^T \sigma_2 \psi_3 - \varphi_3 \psi_1^T \sigma_2
\psi_2)\ .\cr\noalign{\vskip 0.2cm}}$$
The quadratic terms in the superfields are mass terms, and the cubic
contain the renormalizable Yukawa interactions. In addition, they
contain interactions with the auxiliary fields, which lead to
the same mass term for the bosons, and their quartic
renormalizable self-interactions.
Higher order polynomials yield non-renormalizable interactions.

For a real superfield, transforming under supersymmetry like $V$, it is
easy to show that its component along $\vert\theta^T\sigma_2
 \theta\vert^2 $ transforms as a four-divergence. This term is called
the D-term. Thus its space-time integral is a supersymmetric invariant.
By integrating over both $\theta$ and $\theta^\ast$, we can extract the
D-term. Our only example so far is the kinetic part of the Action

$$\int d^4x\int d^2\theta\int d^2\overline\theta \vert
\Phi(x_\mu+\theta^\dagger\sigma_\mu\theta,\theta)\vert^2\ .$$
It has the right dimension: the superfield has dimension one, and the
four Grassmann integral bring dimension two.

The potential part of the Action is given by
$$\int d^4x\int d^2\theta P(\Phi) + c.c.\ ,$$
where the function $P$ is called the superpotential; it depends only on
the chiral superfields, {\bf not} their conjugates. For renormalizable
theories, it is at most cubic

$$P=m_{ij}\Phi_i\Phi_j+\lambda_{ijk}\Phi_i\Phi_j\Phi_k\ .$$
It is straightforward to see that the physical potential is
simply expressed in terms of the superpotential
$$V(\varphi)=\sum_iF^\ast_iF_i^{}=
\sum_i\vert{\partial P(\varphi)\over {\partial\varphi_i}}\vert^2\ ;$$
it is obviously positive definite, which is a general feature of global
supersymmetry.

It is easy to implement internal symmetries: just assume that the whole
superfield transforms as some representation of some internal group. The
kinetic part (only if the invariance is global) is automatically
invariant by summing over all the internal degrees of freedom. The
superpotential may not be invariant, which restricts its form.

The kinetic term has a special global symmetry, called R-symmetry; it is
not an internal symmetry since it does not commute with
supersymmetry. R-symmetry is a global phase symmetry on the
Grassmann variables
$$\theta\to e^{i\beta}\theta
\ ,\qquad \theta^\ast\to e^{-i\beta}\theta^\ast\ .$$
This means that the Grassmann measures transform in the
opposite way
$$d\theta\to e^{-i\beta}d\theta\ ,\qquad d\theta^\ast\to
e^{i\beta}d\theta^\ast\ .$$
The Grassmann integration measure for the kinetic term is invariant. The
most general R-type transformation that leaves the kinetic integrand
invariant is

$$\Phi_i(x_\mu,\theta)\to e^{in_i\beta}\Phi_i(x_\mu,e^{i\beta}\theta)\ .$$
This symmetry is not necessarily shared by the superpotential, unless it
transforms under R as $$P\to e^{2i\beta}P\ ,$$ to match the
transformation of the Grassmann measure. This further restricts the form
of the superpotential.

To see the role of the auxiliary fields, consider two chiral superfields
with only the mass term in their superpotential, $m\Phi_1\Phi_2$.
In the Lagrangian density we find the terms
$$F_1^\ast F_1+F_2^\ast F_2+
\{m(\varphi_1 F_2 +
\varphi_2 F_1 - \psi_1^T \sigma_2 \psi_2)+ {\rm c.c.}\}\ .$$
The equations of motions for the auxiliary fields such as
$$F^*_1=-m\varphi_2\ ,\ etc\ ,$$
allow us to simply rewrite them in terms of the physical fields, with
the result
$$-m^2 \vert\varphi_1\vert^2 -m^2\vert\varphi_2\vert^2 - m \psi_1^T
\sigma_2 \psi_2\ .$$
These are the mass terms for four real scalars and one Dirac fermion of
mass $m$. This leads us to to the mass sum rule

$$\sum_{J=0} m^2 = 2 \sum_{J=1/2} m^2\ ,$$
where we count the number of Weyl fermions (1 Dirac = 2 Weyl).
If we had only one superfield with interaction ${m\over 2} \Phi
\Phi$, the extra term in the Lagrangean would have been simply
$$m(\varphi F - {1\over 2} \psi^T \sigma_2 \psi)\ ,$$
which describes one complex scalar of mass $m$ and one Weyl of
mass $m$; in this case the sum rule
$$\sum_{J=0, 1/2} (2J+1) (-1)^{2J} m_J^2 = m^2 + m^2 - 2m^2=0\ ,$$
is again satisfied.
\bigskip
\noindent Functions of a Chiral Superfield

In the following, we work out certain functions of superfields, which
are of some interest in discussing non-renormalizable supersymmetric
theories. As we have seen, products of chiral superfields are themselves
chiral superfields, so that any special function of a chiral superfield
is defined through its series expansion.

\vskip .2cm
\noindent {\it Logarithm}

Given a chiral superfield
$$\Phi = \varphi(x) + \theta^T\sigma_2 \psi(x) +
{1\over 2}\theta^T\sigma_2 \theta F(x)\ ,$$
we have
$$\eqalign{\ln \Phi &= \ln\{\varphi[1 + \theta^T\sigma_2
\hat\psi (x) + {1\over 2}\theta^T\sigma_2 \theta \hat F(x)]\}\
,\cr\noalign{\vskip 0.2cm}
&= \ln \varphi + \ln[1 + \theta^T\sigma_2 \hat \psi (x)
+{1\over 2}\theta^T\sigma_2 \theta \hat F]\ ,\cr\noalign{\vskip 0.2cm}}$$
where
$$\hat \psi = {\psi\over \varphi}\ ,~~~\hat F = {F\over \varphi}\
.$$
We then use the series expansion of the logarithm to obtain
$$\eqalign{\ln \Phi &= \ln \varphi + (\theta^T\sigma_2
\hat\psi + {1\over 2}\theta^T\sigma_2 \theta \hat F) -
{1\over 2} (\theta^T\sigma_2 \hat \psi +
{1\over 2}\theta^T\sigma_2 \theta \hat F)^2\
,\cr\noalign{\vskip 0.2cm}
&= \ln \varphi + \theta^T\sigma_2 \hat \psi +{1\over 2} \theta^T
\sigma_2 \theta (\hat F + {1\over 2} \hat \psi^T \sigma_2 \hat
\psi)\ ,\cr\noalign{\vskip 0.2cm}}$$
where we have used the Fierz identities and the fact that the
expansion in $\theta$ cuts off after second order.
\vskip .2cm
\noindent {\it Power}

The arbitrary power of a chiral superfield is given by its series
expansion, since

$$\eqalign{\Phi^a &= \varphi^a\{1+\theta^T\sigma_2\hat\psi+
{1\over 2}\theta^T\sigma_2\theta\hat F\}^a\
,\cr\noalign{\vskip 0.2cm}
&=\varphi^a\{1+a\theta^T\sigma_2\hat\psi+{1\over 2}a\theta^T\sigma_2
\theta
\hat F + {1\over 2}a(a-1) (\theta^T\sigma_2 \hat \psi)^2\}]\
,\cr\noalign{\vskip 0.2cm}}$$
which, after a Fierz, yields the exact result.

$$\Phi^a = \varphi^a [1 + a \theta^T\sigma_2 \hat \psi
+{1\over 2}\theta^T\sigma_2 \theta (a \hat F - {a(a-1)\over 2} \hat \psi^T
\sigma_2 \hat \psi)]\ .$$
\vskip 1cm
\centerline{The Real Superfield}
\vskip .5cm
We have already seen how to construct a real superfield out of a chiral
superfield. In general, however, we should be able to build it in terms
of the four real Grassmann variables which describe the Weyl spinor
$\theta$. An elegant way to do this is to rewrite the two-component Weyl
into a four component Majorana spinor. In the Majorana representation
for the Dirac matrices, all four components of a Majorana spinor are
real, so that we are dealing with four real anticommuting degrees of
freedom. The real superfield is  the most general expansion in terms of
the real Majorana spinor $$\Theta = \left ( \theta \atop -\sigma_2
\theta^\ast \right )\ ,$$ shown here in the Weyl representation.

Because they anticommute, the expansion will stop at the fourth order.
Naive counting results in having 4 components to the first order,
${4.3\over 2} = 6$ components at the second, ${4\cdot 3\cdot 2\over
1\cdot 2\cdot 3} = 4$ at the third, and ${4\cdot 3\cdot 2\cdot\over
1\cdot 2\cdot 3\cdot 4} = 1$ component at the fourth.  Hence a real
superfield contains $(1,4,6,4,1)$ degrees of freedom, half commuting,
half anti-commuting. We can form the six quadratic covariants

$$\overline \Theta
\Theta,~~\overline \Theta \gamma_5 \Theta,~~\overline \Theta \gamma_5
\gamma_\mu \Theta\ ,$$ where
the bar denotes the usual Pauli adjoint
$$\overline \Theta= \Theta^\dagger \gamma^0\ .$$
It is easy to check the reality conditions
$$\eqalign{(\overline \Theta \Theta)^\ast &= - \overline
\Theta \Theta\ ,\qquad (\overline \Theta \gamma_5 \Theta)^\ast =
\overline \Theta \gamma_5 \Theta\ ,\cr\noalign{\vskip 0.2cm} (\overline
\Theta \gamma_5 \gamma_\mu \Theta)^\ast &= - \overline \Theta \gamma_5
\gamma_\mu \Theta\ .\cr\noalign{\vskip 0.2cm}}$$
The further identities
$$\overline \Theta \Theta \overline \Theta = - \overline \Theta \gamma_5
\Theta \overline \Theta \gamma_5 = {1\over 4} \overline \Theta \gamma_5
\gamma_\mu \Theta \overline \Theta \gamma_5 \gamma^\mu\ ,$$
$$\overline \Theta \gamma_5 \gamma_\mu \Theta \overline \Theta = -
\overline\Theta
\Theta \overline \Theta \gamma_5 \gamma_\mu\ ,\qquad
\overline \Theta \gamma_5 \gamma^\mu \Theta \overline \Theta
\gamma_5\gamma^\mu \Theta = g^{\mu\nu} (\overline \Theta \Theta)^2\ ,$$
are useful in arriving at the general Lorentz covariant expansion
of a real superfield
$$\eqalign{V(x^\mu,\Theta) &= A(x) + i \overline \Theta
\Psi(x) + i \overline
\Theta \Theta M(x) + \overline \Theta \gamma_5 \Theta
N(x)\cr\noalign{\vskip 0.2cm}
{}~~~~~&+ i \overline \Theta \gamma_5 \gamma^\mu \Theta A_\mu (x) +
\overline
\Theta \Theta \overline \Theta \Lambda(x) + (\overline \Theta
\Theta)^2 D(x)\ .\cr\noalign{\vskip 0.2cm}}$$
In Weyl notation, the same real superfield reads
$$\eqalign{V(x^\mu,\theta,\theta^*)=A(x)
&-i( \theta^T\sigma_2 \psi +
\theta^\dagger \sigma_2 \psi^\ast)\cr\noalign{\vskip 0.2cm}
&-i \theta^T\sigma_2 \theta C -i \theta^\dagger \sigma_2
\theta^\ast C^\ast+ i \theta^\dagger \sigma^\mu \theta A_\mu
\cr\noalign{\vskip 0.2cm}& + \theta^T\sigma_2
\theta \theta^\dagger \sigma_2 \lambda + \theta^\dagger \sigma_2
\theta^\ast \theta^T\sigma_2 \lambda^*
+ \vert\theta^T\sigma_2 \theta\vert^2 D\ ,\cr\noalign{\vskip 0.2cm}}$$
where
$$C(x)=M(x)-iN(x)\ ,$$
and
$$\Psi (x) = \left ( \psi (x) \atop -\sigma_2 \psi^\ast
(x) \right )\ ,\qquad \Lambda = \left ( \lambda \atop -\sigma_2
\lambda^\ast \right )\ .$$
The real superfield also contains a chiral superfield and its conjugate,
made up of the non-canonical fields $A$, $\psi$, and $C$. We can always
write it in the form
$$V(x,\theta,\theta^*)=-i(\Phi(x,\theta)-\Phi^*(x,\theta))+\hat
V(x,\theta,\theta^*)\ ,$$
where
$$\Phi(x,\theta)={1\over 2}(B(x)+iA(x))+\theta^T\sigma_2\psi(x)+
\theta^T\sigma_2\theta C(x)\ .$$

If the real superfield is  dimensionless, the vector
field $A_\mu$ and the Weyl spinor $\lambda$ have the right canonical
dimension to represent a gauge field, and a spinor field. The real
superfield describes the vector supermultiplet we
have encountered in classifying the representations of the super
Poincar\' e group, but with many extra  degrees of freedom, which
happen to fall neatly in  chiral multiplets. This is no accident, since
they in fact  turn out to be gauge artifacts.

For future reference, let us work out some functions of a real
superfield which are useful  in some physical applications.
Starting with
the power of a real superfield, we have
$$V^a = [A(1 + X)]^a\ ,$$
with
$$X = i \overline\Theta \hat\Psi + i \overline\Theta \Theta
\hat M + \overline\Theta \gamma_5 \Theta \hat N + i \overline\Theta
\gamma_5 \gamma^\mu \Theta \hat A_\mu
+ \overline\Theta \Theta \overline\Theta \hat \Lambda + (\overline
\Theta
\Theta)^2 \hat D\ ,$$
where the hat denotes division by $A$.  Then, noting that
$X^5 =
0$, a little bit of algebra gives
$$\eqalign{V^a = A^a [1& + aX + a(a-1) {X^2\over 2!}\cr
& + a(a-1)(a-2)
{X^3\over 3!} + a(a-1)(a-2)(a-3) {X^4\over 4!}]\cr}\ .$$
The Fierz identity shown here for any two Dirac four component
spinors
$$\Psi \overline\Lambda = - {1\over 4} \overline\Lambda \Psi -
{1\over 4}
\gamma_5 \overline\Psi \gamma_5 \Lambda + {1\over 4} \gamma_5
\gamma^\rho
\overline\Psi \gamma_5 \gamma^\rho\Lambda - {1\over 4} \gamma^\rho
\overline\Psi
\gamma_\rho \Lambda + {1\over 2} \sigma_{\mu\nu} \overline\Psi
\sigma^{\mu\nu} \Lambda \ ,$$
is used repeatedly to rewrite the powers of X in terms of the
standard expansion for a real superfield.  We leave it as an
exercise in fierce Fierzing to work out the general formula.
Here we just concentrate on the D-term. The contributions to the
D-term are as follows:
$$\eqalign{X~:&~ (\overline\Theta \Theta)^2 \hat D\
;\cr\noalign{\vskip 0.2cm}
X^2~: &~2i \overline\Theta \hat\Psi \overline\Theta \Theta
\overline\Theta \hat\Lambda + (i \overline\Theta \Theta \hat M +
\overline\Theta
\gamma_5 \Theta \hat N + i \overline\Theta \gamma_5 \gamma^\mu
\Theta
\hat A_\mu)^2\ ,\cr\noalign{\vskip 0.2cm}
&~= (\overline\Theta \Theta)^2 \left \{ - {i\over 2}
\hat{\overline\Lambda}
\hat\Psi -M^2 + N^2 - \hat A_\mu \hat A^\mu \right \}
\ ;\cr\noalign{\vskip 0.2cm}
X^3~: &~-3(i \overline\Theta \Theta \hat M + \overline\Theta
\gamma_5 \Theta \hat N + i \overline\Theta \gamma_5 \gamma^\rho
\Theta
\hat A_\rho)(\overline\Theta \hat\Psi)^2\ ,\cr\noalign{\vskip
0.2cm}
&~= {3\over 4} (\overline\Theta \Theta)^2 (i \hat M
\hat{\overline\Psi}
\hat\Psi - \hat N \hat{\overline\Psi} \gamma_5 \hat\Psi - \hat
A^\rho
\hat{\overline\Psi} \gamma_5 \gamma_\rho \hat\Psi)\
;\cr\noalign{\vskip 0.2cm}
X^4~: &~(i\Theta \hat\Psi)^4\ ,\cr\noalign{\vskip 0.2cm}
&~= {1\over 16} (\overline\Theta \Theta)^2 (\hat{\overline\Psi}
\Psi)^4 [1 + 1
+ g_\mu^\mu] \ .\cr\noalign{\vskip 0.2cm}}$$
Putting it all together, we obtain for the D-term

$$\eqalign{(V^a)_D = &A^a [a \hat D + {a(a-1)\over 2} (-
{i\over
2} \hat{\overline\Lambda} \hat\Psi - \hat M^2 + \hat N^2 - \hat
A_\mu \hat
A^\mu)\cr\noalign{\vskip 0.2cm}
&+ {1\over 8} a(a-1)(a-2) (i \hat M \hat{\overline\Psi} \hat\Psi -
\hat
N \hat{\overline\Psi} \gamma_5 \hat\Psi - \hat A^\rho
\hat{\overline\Psi} \gamma_5 \gamma_\rho \hat\Psi)]\ .
\cr\noalign{\vskip 0.2cm}}$$
You can use these formulae to show that the real superfield, expunged of
its chiral components, satisfies $\hat V^3=0$.

\vskip 1cm
\centerline{The Vector Supermultiplet}
\vskip 1cm
\noindent We have seen that group theory implies the existence of a vector
supermultiplet which generalizes gauge fields to supersymmetry; it
contains (taking the Abelian case for simplicity)

$$\eqalign{&A_\mu(x):~{\rm a~~gauge~field}\cr\noalign{\vskip 0.2cm}
&\lambda(x):~{\rm
a~~Weyl~spinor~(called~the~gaugino)},\cr\noalign{\vskip 0.2cm}
&D(x):~{\rm an~~auxiliary~field}\ .\cr\noalign{\vskip 0.2cm}}$$
The auxiliary field is here to provide the right count between bosonic
and fermionic degreees of freedom. Without using the massless Dirac
equation, the spinor is described by four degrees of freedom. The gauge
field is described by three degrees of freedom, leaving the $D$ to make
up the balance. With the use of the equations of motion, the Weyl field
has two degrees of freedom, and so does the massless gauge field, and
the auxiliary field disappears.

Sometimes the gaugino is called a Majorana fermion, but there should be
no confusion between a Weyl fermion and a Majorana fermion: in
two-component notation they look exactly the same.

The Action
$$S = \int d^4 x [- {1\over 4} F_{\mu\nu} F^{\mu\nu} +
\lambda^\dagger \sigma^\mu \partial_\mu \lambda + {1\over 2}
D^2]\ ,$$
where $F_{\mu\nu} = \partial_\mu A_\nu - \partial_\nu A_\mu$, is
invariant under the following supersymmetry transformations
$$\eqalign{\delta A_\mu &= -i \lambda^\dagger \sigma_\mu \alpha -
i \alpha^\dagger \sigma_\mu \lambda\ ,\cr\noalign{\vskip 0.2cm}
\delta \lambda &= {1\over 2}( D + {i\over 2} \sigma^{\mu\nu}
F_{\mu\nu})\alpha\ ,\cr\noalign{\vskip 0.2cm}
\delta D &= \partial_\mu \lambda^\dagger \sigma^\mu \alpha -
\alpha^\dagger \sigma^\mu \partial_\mu \lambda\ ,\cr\noalign{\vskip
0.2cm}}$$
where
$$\sigma^{\mu\nu} = {1\over 2} (\overline\sigma^\mu \sigma^\nu -
\overline\sigma^\nu \sigma^\mu)\ .$$
Note again that D transforms as a four-divergence, so that the
integral of D is a supersymmetric invariant.  Let us check the
commutation relations of the algebra:
$$\eqalign{\delta_1 \delta_2 D & = \partial_\mu \delta_1
\lambda^\dagger \sigma^\mu \alpha^{}_2 - \alpha_2^\dagger
\sigma^\mu
\partial_\mu \delta_1 \lambda\ ,\cr\noalign{\vskip 0.2cm}
& = {1\over 2}(\alpha_1^\dagger \sigma^\mu \alpha^{}_2
-\alpha_2^\dagger \sigma^\mu
\alpha^{}_1) \partial_\mu D -\cr\noalign{\vskip 0.2cm}
&~~~~~~~~~-{i\over 4}\left((\sigma^{\rho\sigma} \alpha_1)^\dagger
\sigma^\mu \alpha^{}_2
- \alpha_2^\dagger \sigma^\mu
\sigma^{\rho\sigma}
\alpha^{}_1 \right)\partial_\mu F_{\rho\sigma}\ .\cr\noalign{\vskip
0.2cm}}$$
The use of
$$\sigma^{\rho\sigma\dagger} = - \overline\sigma^{\rho\sigma} = -
{1\over 2} (\sigma^\rho \overline\sigma^\sigma - \sigma
\overline\sigma^\rho)\ ,$$
and of the identity
$$\sigma^\mu \sigma^{\rho\tau} = - i \epsilon^{\mu\rho\tau\delta}
\sigma_\delta + g^{\mu\rho} \sigma^\tau - g^{\mu\tau}
\sigma^\rho\ ,$$
and its conjugate, leads us to the equation
$$\eqalign{[\delta_1, \delta_2] D &=  \alpha_1^\dagger
\sigma^\mu \alpha_2 \partial_\mu D\cr\noalign{\vskip 0.2cm}
&~~~+ {i\over 4} \alpha_1^\dagger (\overline\sigma^{\rho\tau}
\sigma^\mu
+ \sigma^\mu \sigma^{\rho\tau}) \alpha_2
\partial_\mu F_{\rho\tau} - (1\leftrightarrow 2)\ ,
\cr\noalign{\vskip 0.2cm}}$$
whence
$$\eqalign{[\delta_1, \delta_2] D = &(\alpha_1^\dagger \sigma^\mu
\alpha^{}_2
- \alpha_2^\dagger \sigma^\mu \alpha^{}_1) \partial_\mu D
\cr\noalign{\vskip 0.2cm}
&~~+ {1\over 2}(\alpha_1^\dagger \sigma_\lambda\alpha^{}_2 - \alpha_2^\dagger
\sigma_\lambda\alpha^{}_1) \epsilon^{\mu\rho\tau\lambda}
\partial_\mu F_{\rho\tau}\ .\cr\noalign{\vskip 0.2cm}}$$
The last term vanishes because of the Bianchi identity.  (What
if it did not?  Any implications for the monopole?)
Similarly, we compute
$$\eqalign{[\delta_1, \delta_2] A_\mu &= -i \delta_1
\lambda^\dagger \sigma_\mu \alpha^{}_2 - i \alpha_2^\dagger
\sigma_\mu \delta_1 \lambda - (1 \leftrightarrow 2)\
,\cr\noalign{\vskip 0.2cm}
&= {1\over 4} \alpha_1^\dagger (\overline \sigma^{\rho\tau}
\sigma_\mu
- \sigma_\mu \sigma^{\rho\tau}) \alpha^{}_2 F_{\rho\tau} - (1
\leftrightarrow 2)\ ,\cr\noalign{\vskip 0.2cm}
&= (\alpha_1^\dagger \sigma^\rho \alpha^{}_2 - \alpha_2^\dagger
\sigma^\rho \alpha^{}_1) F_{\rho\mu}\ ,\cr\noalign{\vskip 0.2cm}}$$
skipping over several algebraic steps.  The right hand side
contains the desired term, namely $\partial_\rho A_\mu$, but it
also contains $-\partial_\mu A_\rho$; clearly it could not be
otherwise from the transformation laws:  their right-hand side is
manifestly gauge invariant, which $\delta A_\mu$ certainly is
not.  Indeed our result can be rewritten in the form
$$[\delta_1, \delta_2] A_\mu = (\alpha_1^\dagger \sigma^\rho
\alpha^{}_2 - \alpha_2^\dagger \sigma^\rho \alpha^{}_1)
\partial_\rho
A_\mu - \partial_\mu \Sigma\ ,$$
where the field dependent gauge function is
$$\Sigma = (\alpha_1^\dagger \sigma^\rho \alpha^{}_2 -
\alpha_2^\dagger \sigma^\rho \alpha^{}_1) A_\rho\ .$$
This shows clearly that a supersymmetry transformation (in this form) is
accompanied by a gauge transformation. It also means that the
description of the gauge multiplet we have just presented is not gauge
invariant, but rather in a specific gauge; this gauge is called the
Wess-Zumino gauge. It is possible to eliminate the gauge transformation
in the commutator of two supersymmetries by introducing extra fields
which are needed for a gauge invariant description. We leave it as an
exercise to derive the full gauge invariant set of fields. These fields
can be neatly assembled in a real superfield, which is not gauge
invariant, but undergoes the transformation

$$V\to V+i(\Xi-\Xi^\ast)\ ,$$
where $\Xi(x,\theta)$ is a chiral superfield. This nicely connects with the
remarks of the previous section. The Wess-Zumino gauge is that for which
the extraneous
components of the real superfield are set to zero ($A=\psi=C=0$).

Verify that the third commutator yields the expected result
$$[\delta_1, \delta_2] \lambda = (\alpha_1^\dagger \sigma^\mu
\alpha_2 - \alpha_2^\dagger \sigma^\mu \alpha_1) \partial_\mu
\lambda(x)\ .$$

Generalization to the non-Abelian case is totally straightforward. The
only difference is that the gaugino and auxiliary fields $\lambda^A(x)$
and $D^A(x)$ now transform covariantly as members of the adjoint
representation.  Thus the ordinary derivative acting on $\lambda^A(x)$
has to be replaced by the covariant derivative

$$({\cal D}_\mu \lambda)^A = \partial_\mu \lambda^A + ig(T^C)^A
_B A_\mu^C \lambda^B\ ,$$
where the representation matrices are expressed in terms of the
structure functions of the algebra through
$$(T^C)^A _{~B} =-if^{~CA} _B\ .$$

The non-Abelian Yang-Mills Lagrangean generalizes to

$$- {1\over 4} G_{\mu\nu}^A G_{}^{A \mu\nu} + \lambda^{\dagger
A} \sigma^\mu ({\cal D}_\mu \lambda)^A + {1\over 2} D^A D^A\ .$$
 In the Wess-Zumino gauge, the fields of the vector supermultiplet
can also be very neatly arranged in a chiral superfield which transforms as
a Weyl spinor under the Lorentz group.  It is given by

$$W^A(x,\theta) = \lambda^A (x) + {1\over 2}\left[D^A (x) + {i\over 2}
\sigma^{\mu\nu} G_{\mu\nu}^A (x) \right]\theta + {1\over 4}\theta^T\sigma_2
\theta \overline \sigma^\mu \sigma_2 \partial_\mu \lambda^{\ast
A} (x)\ ,$$
where we have not shown the spinor index.  Under a gauge
transformation,
this superfield transforms covariantly, as a member of the adjoint
representation. One can also easily show that,
under a supersymmetry transformation, $W^A(x,\theta)$ does indeed
transform as a chiral superfield, that is
$$W^A (x^\mu, \theta) \rightarrow W^A (x^\mu + \alpha^\dagger
\sigma^\mu \theta, \theta + \alpha)\ .$$

This reformulation allows us to easily build invariants out of
products
of this superfield. As for the Wess-Zumino multiplet, invariants
are the F-term of the products of this superfield. This time, we must
take care that Lorentz and gauge invariance is satisfied.
In particular, the Yang-Mills Lagrange density is just
$${\cal L}_{SYM}=\int d^2\theta  (W^{A})^T\sigma_2W^A+{\rm c.c.}\ .$$
One can form another invariant
$${\cal L}_{SST}=i\int d^2\theta ( W^{A})^T\sigma_2W^A+{\rm c.c.}\ ,$$
which is the usual Yang-Mills surface term
$${\cal L}_{SST}=  G^A_{\mu\nu}\tilde
G^{A\mu\nu}-i\partial_\mu(\lambda^\dagger\sigma^\mu\lambda)\ .$$

We can make many other supersymmetric invariants;
for $SU(N)$ with $N>2$, we can consider the gauge adjoint $``$anomaly"
composite
$$\int d^2\theta d^{ABC}( W^A)^T\sigma_2W^B\ .$$
One can even build composites which transform as a self-dual
antisymmetric second rank Lorentz tensor, and member of the adjoint
representation of the gauge group, such as
$$f^{ABC}(W^B)^T\sigma_2\sigma^iW^C\ .$$
Some of these constructions prove useful in the context of
supersymmetric dynamical models.

Finally, we can implement R-symmetry on the gauge supermultiplet,
provided that
$$W\to e^{i\beta}W\ ,$$
which means that the gaugino itself carries one unit of R-symmetry,
and the $D$ and gauge fields have no R-number.
\vskip 1cm
\centerline{Chiral and Vector Supermultiplets in Interaction}
\vskip .5cm
Renormalizability restricts the spin of the fields to be no higher than
one-half. For supersymmetry, it means that the only type of matter that
can couple to the gauge supermultiplet is a collection of chiral
Wess-Zumino multiplets.

Let us first consider the coupling of a Wess-Zumino multiplet to an
Abelian gauge superfield. Consider the Action for a number of chiral
superfields. It is clearly invariant under the global phase
transformations

$$\Phi_a(x,\theta)\to e^{i\eta_a}\Phi_a(x,\theta)\ ,$$
as long as the $\eta_a$ are constants, independent of the coordinates.
Assume that the superpotential is invariant under these transformations.
Then its invariance group is much larger. Indeed, the most general {\it
local} phase transformation on these chiral superfields which leaves the
superpotential invariant is

$$\Phi_a(x,\theta)\to e^{i\eta_a\Xi(x,\theta)}\Phi_a(x,\theta)\ ,$$
where $\Xi(x,\theta)$ is a chiral superfield. The kinetic term,
however is no longer invariant, since

$$\Phi^*_a(y,\theta)\Phi_a(y,\theta)\to
e^{i\eta_a(\Xi(y,\theta)-\Xi^*(y,\theta))
}\Phi^*_a(y,\theta)\Phi_a(y,\theta)\ ,$$
where
$$y_\mu=x_\mu+{1\over 2}\theta^\dagger\sigma_\mu\theta\ .$$
This is analogous to the situation in usual field theory.
To restore invariance, the kinetic term is generalized by adding a real
superfield, which transforms as
$$V\to V-i(\Xi-\Xi^*)\ .$$
The change of the argument translates in a redefinition of $\Lambda(x)$ and
$D(x)$ in the real superfield, and does not affect the
counting of the number of degrees of freedom.

The new kinetic Action is just
$$\int d^4x\int d^2\theta d^2\bar\theta
\sum_a\Phi^*_a(y,\theta)e^{\eta_aV(y,\theta,\theta^*)} \Phi_a(y,\theta) \ .$$
In the Wess-Zumino gauge, this expression reduces to

$$\eqalign{{\cal L} = &- {1\over 4} F_{\mu\nu} F^{\mu\nu} +
\lambda^\dagger \sigma^\mu \partial_\mu \lambda + {1\over 2}
D^2\cr\noalign{\vskip 0.2cm}
&+ ({\cal D}_\mu \varphi)^\ast ({\cal D}^\mu \varphi)^\ast +
\psi^\dagger \sigma^\mu {\cal D}_\mu \psi + F^\ast
F\cr\noalign{\vskip 0.2cm}
&+ gD \varphi^\ast \varphi - 2g \lambda^T\sigma_2
\psi \varphi^\ast +2g\lambda^\dagger\sigma_2\psi^\ast \varphi\ ,
\cr\noalign{\vskip 0.2cm}}$$
with the gauge covariant derivatives
$${\cal D}_\mu \varphi = (\partial_\mu + i gA_\mu)\varphi\ ;~
{\cal
D}_\mu \psi = (\partial_\mu + igA_\mu)\psi\ .$$
The last line gives new interactions, over the usual construction of
gauge invariant theories, with derivatives replaced by covariant
derivatives. The reason is that the new interaction terms created in
this way, all proportional to the charge, are not supersymmetric
invariants. The extra terms restore invariance under supersymmetry.
However it is a bit tricky to check the invariance because we are in the
Wess-Zumino gauge. This entails changes in the transformation properties
of the fields of order g. Let us give some examples.

Consider the variation of the interaction of the fermion current
with the gauge potential. We find
$$\delta\left(ig\psi^\dagger \sigma^\mu \psi A_\mu\right)
=ig\psi^\dagger \sigma^\mu \alpha F A_\mu
+ g \psi^\dagger\sigma^\mu \psi (\lambda^\dagger \sigma_\mu \alpha +
\alpha^\dagger \sigma_\mu \lambda)\ .$$
To offset the last term we need the variation
$$-2g\lambda^T\sigma_2
\psi \delta  \varphi^\ast =
-g\alpha^\dagger\sigma_\mu  \lambda \psi^\dagger\sigma^\mu  \psi \ .$$
By the same token, the variation
$$-2g\delta\lambda^T\sigma_2
\psi \varphi^\ast =-g D\alpha^T\sigma_2
\psi \varphi^\ast  +\cdots\ ,$$
is compensated by
$$gD\varphi^\ast\delta\varphi=gD\varphi^\ast
\alpha^T\sigma_2 \psi\ .$$
This procedure goes on {\it ad nauseam}. The alert student may have
notice the presence of a term proportional to F. The only way to
compensate for it is to add a term in the variation of F itself. The
extra variation
$$\delta_{WZ}F^*=-ig\psi^\dagger\sigma^\mu\alpha A_\mu\ ,$$
does the job.  Its effect is to replace the derivative by the covariant
derivative in the transformation law, which we do for all of them. Even
then we are not finished: we still have one stray
term proportional to F. Indeed we have
$$-2g\lambda^T\sigma_2
\delta\psi \varphi^\ast =-2gF\lambda^T\sigma_2
\alpha  \varphi^\ast  +\cdots\ ,$$
which can only cancelled by adding a term in the variation
of F, yielding
$$\delta_{WZ}F^*=-ig\psi^\dagger\sigma^\mu\alpha A_\mu
-2g\alpha^\dagger\sigma_2\lambda^*\varphi\ .$$
You have my word that it is the last change, but to the non-believer, I
leave the full verification of the modified supersymmetric algebra in
the Wess-Zumino gauge as an exercise during half-time.

This Lagrangian is of course does not lead to a satisfactory quantum
theory because of the ABJ anomaly associated with the U(1); it can be
cancelled by introducing another chiral superfield with opposite charge.
Then the extra terms beyond the covariant derivatives read

$$gD(\varphi_1^\dagger \varphi^{}_1 -
\varphi_2^\dagger \varphi^{}_2)-\left(  2g\lambda^T\sigma_2
(\psi^{}_1 \varphi_1^\ast -\psi^{}_2
\varphi_2^\ast)+~{\rm c.c.}\right)\ .$$
{}From the equations of motion, the value of the auxiliary field is
$$D = -g( \varphi_1^\dagger \varphi^{}_1 - \varphi_2^\dagger
\varphi^{}_2)\ ,$$
yielding the potential
$$V={g^2\over 2}( \varphi_1^\dagger \varphi^{}_1 - \varphi_2^\dagger
\varphi^{}_2)^2\ .$$

Generalization to the non-Abelian case is straightforward. We
merely quote the results for a chiral matter superfield transforming as
a representation $r$ of the gauge group.
The derivatives on the matter fields $\psi_a$ and
$\varphi_a$ are replaced by the covariant derivatives
$${\cal D}_\mu = \partial_\mu + ig {\bf T}^B A_\mu^B\ ,$$
where ${\bf T}^B$ represent the gauge algebra in the representation
of the chiral superfield.
The auxiliary fields $D^A(x)$ now couple through the term
$$gD^A \varphi^{\dagger a} (T^A)_a^{~b}\varphi_b\ ,$$
and the gauginos by the terms
$$-2g\varphi^{\dagger a} (T^A)_a ^{~b} \psi_b^T\sigma_2 \lambda^A +
2g\lambda^{A\dagger}\sigma_2\psi^{\ast}_a (T^A)_a^{~b} \varphi^a\ ,$$
where we have shown the internal group indices (but not the spinor
indices).

To conclude this section, we note that the gauge coupling preserves
R-symmetry, irrespective of the R-value of the chiral superfield. Having
assembled all the pieces necessary for the generalization of
the $N=0$ standard model to $N=1$, we are ready for its description.

\vskip 1cm
\centerline{\bf III-) THE SUPERSYMMETRIC STANDARD MODEL}
\vskip .5cm
The $N=0$ standard model is easily made supersymmetric. Its
Weyl fermions are put in chiral Wess-Zumino multiplets, its gauge
bosons now form vector supermultiplets, and the Higgs boson is part
of a chiral multiplet. We note that an odd number of Weyl fermions
cannot be implemented if there is more than one supersymmetry, which is
the reason we focus on $N=1$.

The alert among you has noticed that the left-handed lepton doublets and
the Higgs doublet have the same gauged electroweak quantum numbers,
although the lepton doublets have one unit of lepton number while the
Higgs has none. Also that there is only one Higgs doublet and three
lepton doublets. Can we build a model where the Higgs is the
superpartner of a lepton doublet, using R symmetry as a compensator for
lepton number? The student is encouraged to try models in this
direction.

The most $``$economical'' way of introducing N=1
supersymmetry is
to associate with every spin 1/2 particle a chiral superfield,
$$L \rightarrow \Phi_L\ ,~~~{\bf Q} \rightarrow \Phi_{\bf Q}\
,~~~\overline {\bf u}\rightarrow \Phi_{\overline {\bf u}}\
,~~~\overline
{\bf d} \rightarrow\Phi_{\overline {\bf d}}\ ,~~~\overline e
\rightarrow
\Phi_{\overline e}\ ;$$
this procedure associates to each fermion a scalar partner of sfermions
with the same electroweak quantum number
$$\eqalign{\Phi^{}_L ~:&~ \left ( \nu_L \atop e_L \right ) ~{\rm
and}~
\left ( \tilde\nu_L \atop \tilde e_L \right )~~({\rm slepton})\ ,\cr
\Phi_{\overline e}~:&~\overline e_L ~{\rm and}~ \tilde
e_R^\ast~~({\rm antislepton})\ ,\cr
\Phi^{}_{\bf Q}~:&~ \left ( {\bf u}^{}_L \atop {\bf d}^{}_L
\right ) ~{\rm and}~ \left ( \tilde
{\bf u}^{}_L \atop \tilde {\bf d}^{}_L \right )~~({\rm squark})\ ,\cr
\Phi_{\overline {\bf u}}~:&~\overline {\bf u}^{}_L ~{\rm and}~
\tilde
{\bf u}_R^\ast~~({\rm antisquark})\ ,\cr
\Phi_{\overline {\bf d}}~:&~\overline {\bf d}^{}_L ~{\rm and}~
\tilde
{\bf d}_R^\ast~~({\rm antisquark})\ .\cr}$$

The Higgs doublet of the Standard Model is interpreted as the scalar
component of a new chiral superfield. This introduces a left-handed
doublet of Weyl fermions, the Higgsinos. However, we cannot stop here,
because this Higgsino doublet makes the hypercharge anomalous in two
different ways. One type of anomaly is the triangle anomaly; the second
is Witten's global anomaly, which says that any theory with an odd
number of half-integer spin representations of $SU(2)$, path-integrates
to zero. Thus we had better do something about it. Both problems are
solved by postulating the existence of another doublet of Higgsinos,
which is the vector-like completion of the first  with opposite
hypercharge (There are ways to chirally cancel anomalies, but they lead
to much more complicated theories, which do not concern us here). We
thus have two chiral superfields in the $N=1$ model:

$$\eqalign{\Phi_{H_d}~:& \left ( \varphi^0 \atop \varphi^- \right
)\ {\rm
and}~\left (
\tilde \varphi_L^0 \atop \tilde \varphi_L^- \right
)~~({\rm Higgsino})\ ,\cr
\Phi_{H_u}~:&  \left ( \varphi^+ \atop \varphi^0 \right
)\ {\rm and}~\left (\tilde\varphi^+ \atop \tilde \varphi^0 \right
)~~({\rm Higgsino})\ .\cr}$$
It is amusing that we come to
the same conclusion from phenomenology: with
only one Higgs superfield, we cannot give masses to both charge 2/3 and
charge -1/3, -1 fermions. Recall that this is possible in the $N=0$
standard model by using the conjugate of the Higgs field in the
coupling. We have seen that supersymmetry-invariant couplings in the
superpotential are analytic functions of the superfields, and do not
involve their conjugates. Hence we need another Higgs doublet of
opposite hypercharge, which is exactly the conclusion reached from
anomaly considerations.

The Yukawa interactions of the standard model are extracted from the
superpotential

$$ {\bf Y}_{ij}^u
\Phi_{\bf Q}^{iT} \Phi_{\overline u}^j \tau_2 \Phi^{}_{H_u} +
{\bf Y}_{ij}^d
\Phi_{\bf Q}^{iT} \Phi_{\overline {\bf d}}^j \tau_2 \Phi^{}_{H_d}
+ {\bf Y}_{ij}^\ell \Phi_L^i \Phi_{\overline e}^j \tau_2
\Phi^{}_{H_d}\ ,$$
The indices $i,j=1,2,3$ label the three chiral families. This
cubic superpotential contains no mass term for the Higgs fields, and
has also many global symmetries, some nefarious to phenomenology.

The reduction of the flavor Yukawa matrices ${\bf Y}_{ij}^{}$ proceeds
as in the $N=0$ model. Without loss of generality, we can bring the
lepton Yukawa to diagonal form,
$${\bf Y}_{ij}^\ell \rightarrow {\bf Y}_{ii}^\ell\ .$$
We diagonalize the down Yukawa, setting
$${\bf Y}^d = {\bf U}_d^T {\bf M}_d {\bf V}^{}_d\ ,\qquad({\bf
M}_d~{\rm
diagonal})\ ,$$
and rewriting the Lagrangean in terms of the superfields
$$\Phi_{\overline {\bf d}}^{\prime i} = ({\bf V}^{}_d
\Phi^{}_{\overline {\bf d}})^i\
,~~~\Phi_{\bf Q}^{\prime i} = ({\bf U}_d \Phi^{}_{\bf Q})^i\ .$$
The same reduction of the up quarks Yukawa matrix yields
$${\bf Y}^u = {\bf U}_u^T {\bf M}_u {\bf V}^{}_u\ ,~~~~({\bf
M}_u~{\rm
diagonal})\ ,$$
while redefining
$$\Phi_{\overline u}^{\prime i} = ({\bf V}^{}_u \Phi_{\overline
u})^i\ .$$
The ${\bf V}_u$ matrix disappears from the Lagrangean, and we have no
further freedom for $\Phi_{\bf Q}$.  Thus the most we can do (after
dropping the primes) is

$${\bf Y}_{ii}^\ell \Phi_L^{Ti} \Phi_{\overline e}^i \tau_2
\Phi^{}_{H_d} + {\bf M}^{ii}_d\Phi_{\bf Q}^{Ti} \Phi_{\overline {\bf
d}}^i \tau_2 \Phi_{H_d} + \Phi_{\bf Q}^{Ti}  ({\cal U}^T)_i^j {\bf
M}^{jj}_u \Phi^j_{\overline u} \tau_2 \Phi^{}_{H_u}\ ,$$

with the flavor mixing matrix
$$\hat {\cal U}^T = {\bf U}_d^T {\bf U}^{}_u\ ;$$
it reduces to the CKM matrix, after Iwasawa decomposition, to expunge
extraneous phases. Thus, the Yukawa couplings of the $N=0$ and $N=1$
models are exactly the same, if one allows for two Higgs of opposite
hypercharge.

The global phase symmetries of this superpotential are easy to identify.
In the lepton sector, we still have conservation of the relative lepton
numbers. In the quark sector, no distinction between families is
allowed, since  the CKM matrix is different from one. Global
transformations on the superfields appear as

$$\Phi_f \rightarrow e^{in_f \eta} \Phi_f\ ,$$
where $f$ denotes the species: $L$, ${\overline e}$, $\overline{\bf u}$,
$\overline {\bf d}$, or ${\bf Q}$.  The transformations which preserve
supersymmetry obey the relations

$$\eqalign{n^{}_{L_i} + n^{}_{\overline e_i} + n^{}_{H_d} &= 0
\ ,~~~~~~~i =
e,\mu,\tau\ ,\cr
n^{}_Q + n^{}_{\overline u} + n^{}_{H_u} &= 0\ ,~~~~~~~{\rm
any~flavor}\ , \cr
n^{}_Q + n^{}_{\overline {\bf d}} + n^{}_{H_d} &= 0\ ,~~~~~~~{\rm
any~flavor}\ .\cr}$$
With only one family, there are  seven fields, with seven independent
phases, obeying three relations from the couplings, leaving four
independent symmetries; these are easily identified to be
\vskip .2cm
\settabs 8\columns
\+&$n_L$&$n_{\overline e}$&$n_Q$&$n_{\overline u}$&$n_{\overline
{\bf d}}$&$n_{H_u}$&$n_{H_d}$\cr
\+&&&&&&&\cr
\+L&1&-1&0&0&0&0&0\cr
\+B&0&0&1/3&-1/3&-1/3&0&0\cr
\+Y&-1&2&1/3&-4/3&2/3&1&-1\cr
\+PQ&-1/2&-1/2&-1/2&-1/2&-1/2&1&1\cr
\vskip .2cm
They are: two global symmetries, total lepton number (L), and baryon
number (B), one local symmetry, hypercharge (Y), and the chiral
Peccei-Quinn (PQ) symmetry. With three families, there are also two
conserved relative lepton numbers, $L_e-L_\mu$ and $L_\mu-L_\tau$.  Of
these, only the Peccei-Quinn symmetry does not occur in the standard
model.

A special feature of supersymmetric theory is the global R-symmetry,
under which

$$\eqalign{\theta &\rightarrow e^{i\eta} \theta\cr
\Phi &\rightarrow e^{i2\eta/3}
\Phi~~~~~~~~~{\rm for~all~chiral~matter~superfields}\ .\cr}$$
The chiral spinor superfields that contain the
gauge bosons transform as well
$$W^a(x,\theta)\rightarrow e^{i\eta}W^a(x,\theta)\ ,$$
(shown here for $SU(2)_W$ only)
so that the gauginos also transform under R
$$\lambda(x)\to e^{i\eta}\lambda(x)\ .$$
This symmetry requires massless gauginos since their Majorana masses
have two units of R. However, gauginos must be massive, if supersymmetry
is to describe the real world, and R symmetry must be broken. We note in
passing that R-symmetry is anomalous but this does not mean that the
gauginos acquire arbitrary masses since they are tied by supersymmetry
to the vector particles whose masses are set by gauge invariance.

The PQ symmetry also causes a problem since it is carried only by fields
that transform as weak isospinors. It is well-known that this leads to a
weakly coupled axion with large mass, a possibility that is
experimentally ruled out. Hence this symmetry must be broken as well.

This embarassment of symmetries is somewhat alleviated when it is
realized that with the minimal set of fields of the $N=1$ model, we can
add to the superpotential the so-called $\mu$-term

$$\mu \Phi_{Hu}^T \tau_2 \Phi^{}_{H_d}\ ,$$
without violating supersymmetry. It introduces in the model a mass term
$\mu$, but breaks both PQ and R symmetries, leaving the
linear combination
$$R'= R+{1\over 3} PQ \ ,$$
invariant, since it has $\Delta PQ= 2$, and $\Delta R = -2/3$.

This symmetry still keeps the gauginos massless. It is anomalous, and
explicitly broken by QCD. Both Higgs superfields and the gauginos carry
one unit of $R^\prime$, while the other chiral superfields carry half a
unit. When $H_u$ and $H_d$ get their electroweak breaking values, it is
spontaneously broken, leading to the unacceptable visible axion model.
The terms we have introduced in the superpotential define the {\it
minimal} $N=1$ standard model.

With the chiral superfields of the standard model, we can add to the
superpotential other renormalizable terms which are invariant under both
supersymmetry and the gauge groups $SU(3)\times SU(2)\times U(1)$. They
are, suppressing all family indices,

$$\Phi_{\overline{\bf
d}}\Phi_{\overline{\bf d}}\Phi_{\overline{\bf u}}\ ;\qquad
\Phi_{\bf
Q}\Phi_{\overline{\bf d}}\Phi^{}_{L}\ ;\qquad
\Phi^{}_{L}\Phi^{}_{L}\Phi_{\overline e}\ ,\qquad \Phi_{L}\Phi_{H_u}\ .$$
The first term violates quark number by three units, and the
others lepton number by one unit. These new terms are allowed by
supersymmetry, but in view of the the excellent experimental limits on
both  baryon and lepton numbers, they should appear with tiny
coefficients, if at all. All violate $R^\prime$ symmetry ({\it mod} 4),
since they have $R^\prime=3/2$,which does not leave any discrete remnant.
In addition, baryon number is broken only ({\it mod} 3), leaving behind
the discrete group $Z_3$. There is also a parity under which all weak
doublets are odd, all singlets even, but this is a consequence of
invariance under the weak $SU(2)$.

If $N=1$ supersymmetric models are to describe the real world, they must
include mechanisms which break supersymmetry, break the electroweak
symmetry, and break $R^\prime$ symmetry.

First we remark that breaking of supersymmetry necessarily generates a
mass for the gauginos, in order to create a mass gap with the massless
gauge bosons. This automatically breaks $R^\prime $ symmetry, but in a
specific way, leaving behind a discrete symmetry. In the minimal $N=1$
standard model this symmetry is $Z_4$, under which we have

$$\Phi_{f}\to i\Phi_{f} \ ;\qquad \Phi_{H_{u,d}}\to -\Phi_{H_{u,d}} \ .$$
In addition, the gauginos are odd
under this symmetry. Since it is an R symmetry, particles and their
superpartners do not have the same multiplicative quantum number.
The
Higgs scalar doublets are odd under this
4-fold symmetry. It would seem that electroweak breaking would breat
it down to $Z_2$, creating potential domain wall problems, but
some of this symmetry can be expressed in terms of
hypercharge, baryon number and lepton number, which means that the only
extra symmetry is $Z_2$, which is R-parity.
R parity is an exact symmetry of the minimal $N=1$ standard
model. It is easy to see that all quarks, leptons, and Higgs bosons are
even under R-parity; all their superpartners are odd.

It has the important consequence that superpartners can only decay
into an odd number of lighter superpartners. Thus, the lightest of
these, the lightest superpartner (LSP) must be stable. While two heavy
to have been produced in the laboratory, many believe that the LSP
pervades the universe as a stable remnant of the cosmological soup; it
might just be what dark matter is made of.

In the non-minimal model, there is no R-parity, and thus no stable
particle, although in some models the LSP could be long-lived.

As we have emphasized, like turtles which carry their own houses,
supersymmetric theories contain their own potential. Thus it is natural
to ask if the potential in the $N=1$ standard model is capable of the
heroic deed of breaking any of the above symmetries. For that purpose,
we must first discuss supersymmetry breaking.
\vskip 1cm
\centerline{\bf IV-) SUPERSYMMETRY BREAKING}
\vskip .5cm
It is time to understand how to break supersymmetry. We disregard hard
breaking, since one of the rationale for supersymmetry is to tame
quantum corrections. We may then consider two different types of
breaking which do not alter the ultraviolet properties of the theory.

One is soft breaking, with the symmetry broken by adding to the theory
terms of dimension 2 and 3.  Intuitively, they do not affect the
theory in the limit where all masses are taken to zero, relative to the
scale of interest. The most direct way is to
give the superpartners of the massless chiral fermions a mass. This can
be done without breaking electroweak symmetry. Also we can give each
Higgs doublet a supersymmetry-breaking mass,and finally we can put in
gaugino mass terms. This clearly splits the mass degeneracy between the
particles within a supermultiplet. This is exactly like adding the
effect of the quark masses in the chiral Lagrangian. Finally, there are
additional possible terms of dimension three between sparticles and
Higgs particles as well. Of these terms, the gaugino Majorana masses,
$$M_i\lambda_i^T\sigma_2\lambda_i\ ,$$
break $R^\prime$ by two units,
leaving the minimal model with an unbroken discrete subgroup, R-parity.
Soft breaking is not fundamental, rather an effective manisfestation of
symmetry breaking.

Another way is spontaneous breaking of supersymmetry, which
we now discuss. A symmetry is spontaneously broken if the field
configuration which yields minimum energy no longer sustains the
transformation under that symmetry. Let us remind ourselves how it works
for a garden variety symmetry. The simplest is when the order parameter
is a complex field $\varphi (x)$ with dynamics invariant under the
following transformation

$$\delta \varphi(x) = e^{i \beta} \varphi(x)\ .$$
Now suppose that in the lowest energy configuration, this field has a
constant value
$$<\varphi (x)>_0 = v \ .$$
Expanding $\varphi (x)$ away from this vacuum configuration, setting

$$\varphi (x) = e^{i\eta(x)}(v + \rho(x))\ ,$$
we find that under the transformation, the angle $\eta(x)$ undergoes a
simple shift
$$\eta(x)\to \eta(x)+\delta\ ,$$
meaning that the dynamics is invariant under that shift. Geometrically,
this variable is the angle which parametrizes the closed line of minima.
The dynamical variable associated with this angle is identified with the
massless Nambu-Goldstone boson, $\zeta(x)$, divided by the vacuum value.
It couples to the rest of the physical system as

$${\cal L}_{NG}={1\over v}\zeta(x)\partial_\mu J^\mu\ ,$$
where $J_\mu(x)$ is the Noether current of the broken symmetry. Clearly,
a constant shift in $\zeta$ generates a surface term and leaves the
Action invariant.

Let us apply this acquired wisdom to the supersymmetric case, starting
with the chiral superfield. In a constant field
configuration, the supersymmetry algebra reads

$$\eqalign{\delta \varphi_0 &= \alpha^T\sigma_2
\psi_0\ ,\cr\noalign{\vskip 0.2cm}
\delta \psi_0 &=  \alpha F_0\ ,\cr\noalign{\vskip 0.2cm}
\delta F_0 &= 0\ .\cr\noalign{\vskip 0.2cm}}$$
Any non-zero value of $\psi_0$ breaks both supersymmetry and
Lorentz invariance. Since we are only interested in
Lorentz-invariant vacua, we set $\psi_0 = 0$, obtaining the only Lorentz
invariant possibility

$$\delta \varphi_0 = 0,~~\delta \psi_0 = \alpha
F_0,~~\delta F_0 = 0\ ;$$
with $\varphi_0 \not= 0$ and $F_0 \not= 0$.
The only way this
configuration can break the supersymmetry is to require that
$$F_0 \not= 0~:~~~~~~{\rm broken~supersymmetry}\ .$$
Since $F$ is a function of the scalar fields, it means that some
$\varphi_0 \not= 0$. It must be noted that when $F_0 = 0$, and
$\varphi_0 \not= 0$, any internal symmetry carried by $\varphi_0$ is
broken. This fits nicely with our earlier remarks because a non-zero
value for $F$ gives the potential a positive minimum.

When $F_0 \not= 0$, the chiral fermion
shifts under supersymmetry:  it is
the Nambu-Goldstone fermion associated with the breakdown
of supersymmetry, as expected,  since the broken symmetric is fermionic.
It often goes under the name Goldstino, although I would prefer, for
historical reasons, to call it Nambino.

A similar analysis carries to the vector multiplet. There,
the only vacuum configuration which does not break Lorentz invariance,
is that where $A_\mu$ and $\lambda$ vanish in the vacuum, for which we
have

$$\delta A_0^\mu = 0\ , ~\delta\lambda_0 = \alpha D_0\ ,~\delta
D_0 = 0\ .$$
It is clear that the only way to break supersymmetry is to give
$D_0$ a vacuum value, and in this case, it is the gaugino $\lambda$
that plays the role of the Nambino (Goldstino).

Thus, as long as we have chiral and vector superfields, the spontaneous
breakdown of supersymmetry comes about when the dynamics is such that
either $F$ or $D$ is non-zero in the vacuum. Another way of arriving at
the same conclusion is to note that the potential from these theories is
given by

$$V = F_i^\ast F_i + {1\over 2} D^2\ ,$$
when $F_i$ and $D$ take on their values obtained from the
equations of motion.  Since $V$ is the sum of positive definite
quantities it never becomes negative and if supersymmetry is
spontaneously broken, its value at minimum is non-zero.

It is possible to formulate a general argument based on the fundamental
anticommutation relations. In theories with exact supersymmetry, the
vacuum state is annihilated by the generators of supersymmetry. However,
the square of the same supersymmetry generators is nothing but the
energy: the energy of the  supersymmetric ground state is necessarily
zero. Since it is also the state of lowest energy, it follows that the
potential is necessarily positive definite. This is what we have just
seen above.

Now suppose that supersymmetry is spontaneously broken. This requires
that the action of supersymmetry on the vacuum is not zero, and
therefore that the vacuum energy be positive. Comparing with the form of
the potential, this can happen only if $F$ and/or $D$ is non-zero.

We can now examine the potential of the minimal $N=1$ standard model.
The F terms couple in the following way

$$\eqalign{&F_{H_u}^T\tau_2  (\mu H^{}_d - \tilde u_R^\ast
{\bf M}^{}_u {\cal U} \tilde Q^{}_L)\cr
+ &F_{H_d}^T\tau_2  (-\mu H^{}_u - \tilde e_R^\ast
{\bf M}_e\tilde L^{}_L - \tilde d_R^\ast {\bf M}_d^{} \tilde {\bf
Q}^{}_L)\cr
+ &F_L^T ({\bf M}^{}_e\tilde e_R^\ast \tau_2 H^{}_d) + (\tilde L^T
{\bf M}^{}_e \tau_2 H^{}_d)F^{}_{\overline e}\cr
+ &F_{\bf Q}^T \tau_2 ({\bf M}_d^{} \tilde d^{}_R H^{}_d + {\cal
U}^T
{\bf M}_u^{}
\tilde u_R^\ast H^{}_u)\cr
+ &(\tilde {\bf Q}_L^T {\cal U}^T {\bf M}^{}_u \tau_2 H^{}_u)
F^{}_{\overline u} + \tilde {\bf Q}^T {\bf M}^{}_d \tau_2 H^{}_d
F^{}_{\overline
d}\ .\cr}$$
The potential coming from these terms is just the sum of the absolute
values squared of the terms which multiply each F. The expression that
results is pretty complicated, but it is not over, as we still have to
get the contribution from the D-terms.  There are three types of
D-terms, corresponding to each of the gauge groups

$$\eqalign{U(1): D &= {1\over 2} g_1 [- \tilde L_{Li}^\dagger
\tilde L^{}_{Li} + 2 \tilde e_{Ri}^\ast \tilde e^{}_{Ri} + {1\over
3}
\tilde {\bf Q}_{Li}^\dagger \tilde {\bf Q}^{}_{Li} + {4\over 3}
\tilde
u_{Ri}^\dagger \tilde u^{}_{Ri} - {2\over 3} \tilde d_{Ri}^\dagger
\tilde d^{}_{Ri}\cr
&~~~~~~~~~~~+ H_u^\dagger H^{}_u - H_d^\dagger H^{}_d]\cr
SU(2): D^a &= g_2 [\tilde L_{Li}^\dagger {\tau^a\over 2} \tilde
L^{}_{Li} + \tilde {\bf Q}_{Li}^\dagger {\tau^a\over 2} \tilde {\bf
Q}^{}_{Li}
+
H_u^\dagger {\tau^a\over 2} H^{}_u + H_d^\dagger {\tau^a\over 2}
H^{}_d]\ ,\cr
SU(3): D^A &= g_3 [\tilde {\bf Q}_{Li}^\dagger {\lambda^A\over 2}
\tilde {\bf Q}^{}_{Li} + \tilde u_{Ri}^\dagger {\lambda^A\over 2}
\tilde
u^{}_{Ri} + \tilde d_{Ri}^\dagger {\lambda^A\over 2} \tilde
d^{}_{Ri}]\ ,\cr}$$
giving to the potential the contribution

$$(D^2 + D^a D^a + D^A D^A)\ .$$
We parenthetically remark that without the $\mu$ term, this potential is
purely quartic. The $\mu$ term gives an equal mass to the Higgs and the
Higgsinos, and also creates cubic couplings among the Higgs and sleptons
and squarks.

It would be too much to hope for this potential to break both
electroweak and supersymmetry. Since it is the sum of squares, its
minimum, if allowed, occurs when all the auxiliary fields are set to
zero. Clearly, there is one solution when all the fields are set to
zero. This solution breaks no symmetry, and thus has the lowest energy.
So, the best we get from this potential is to ask if there are other
degenerate vacua with desirable electroweak breaking features.

To see if it is possible, we work out the value of the potential at the
required electroweak breaking field configuration. We set

$$H_u = {v_u\over {\sqrt 2}} \left ( 0 \atop 1 \right )~~~~~~~
H_d = {v_d\over {\sqrt 2}} \left ( 1\atop 0 \right )\ ,$$
and evaluate the F and D terms, all other fields being zero (unless we
want L-violation as the only charge conserving possibility is $\tilde
\nu_L \not= 0$ which violates L spontaneously).  From

$$\eqalign{H_u^\dagger {\tau^3\over 2} H^{}_u &= - {\vert
v_u\vert^2\over 4}\ ,~~~~~~~~H_d^\dagger {\tau^3\over 2} H^{}_d
={\vert v_d\vert^2\over 4}\ ,\cr
H_u^\dagger H^{}_u &= {\vert v_u\vert^2\over 2}\ ,
{}~~~~~~~~~~~~H_d^\dagger H^{}_d = {\vert v_d\vert^2\over 2}\ ,\cr}$$
we find that, in the electroweak vacuum, there are non vanishing D and F
fields, namely

$$\eqalign{D &= {1\over 4} g_1 [\vert v_u\vert^2 - \vert
v_d\vert^2]\ ,\cr
D^3 &= -{1\over 4} g_2 [\vert v_u\vert^2 - \vert v_d\vert^2]
\ ,\cr
F_{H_u} &= \mu \tau_2 {v_d\over {\sqrt 2}} \left ( 1\atop 0
\right )\ ,\cr F_{H_d} &= - \mu \tau_2 {v_u\over {\sqrt 2}} \left
(0\atop 1 \right )\ .\cr}$$
We conclude that the electroweak vacuum configuration is not a minimum
of this potential. Thus more physics has to be added to this minimal
model, in the form of mechanisms for both supersymmetry and
electroweak symmetry breakings.

There are many models of spontaneous supersymmetry breaking in the
context of renormalizable theories. Models with F breaking were
first investigated by O'Raifeartaigh. Their general feature is that they have
a global R-like symmetry, and they preserve the sum rule

$${\rm Str}{\cal M}^2\equiv \sum_{J=0,1/2}(-1)^{2J}(2J+1)m^2_J=0\ .$$
The other type of breaking is D-breaking, or Fayet-Iliopoulos breaking.
It requires a local $U(1)$ symmetry, to allow for a gauge
singlet term linear in the D associated with the gauge supermultiplet to the
Lagrangian. The sum rule is modified to read

$${\rm Str}{\cal M}^2=gD{\rm Tr} \sum_iq_i\ ,$$
where $q_i$ are the charges of the Weyl fermions. If the anomaly of this
$U(1)$ is cancelled in a vector-like way, the right-hand side is zero.

In both cases, these sum rules cause phenomenological problems, although
they can be modified by quantum corrections. For that reason, such
models have not proved easy to implement. It is a good thing that these
sum rules are modified when supersymmetry is extended to supergravity.
Supergravity, of which I say little, is the theory of local
supersymmetry. It generalizes gravity, and must be present in a
supersymmetric world. It also has the advantage of eating the massless
Nambino, when the spin 3/2 gravitino gets a mass from supersymmetry
breaking.

Let us conclude this short survey with a discussion of dynamical
supersymmetry breaking. Explicit renormalizable models of F- and D-type
breakings of global supersymmetry have appeared in the literature. It
has proven much more difficult to produce models of dynamical breaking
of global supersymmetry.

By multiplying two chiral superfields, we obtain a composite superfield
with components

$$\varphi_1 \varphi_2\ ,~ \psi_1 \varphi_2 + \psi_2 \varphi_1 \
,~\varphi_1 F_2 + \varphi_2 F_1 - \psi_1^T \sigma_2 \psi_2\ .$$
It would appear that its  F-term could acquire a non-zero  vacuum
value, if the fermions were subject to a strong force,  which, in
analogy with chiral symmetry,  would cause a condensate like $\psi_1^T
\sigma_2 \psi_2$ to form. This would break supersymmetry.  In the
Standard Model, such condensates occur as a result of QCD.  Hence if
these fields were like quarks and antiquarks, supersymmetry could be
broken dynamically when quarks condense.

On the other hand, we do not expect a gaugino condensate to break global
supersymmetry, since it is not part of an F component of a chiral
composite. Indeed the gaugino condensate appears as the scalar term of
$(W^A)^T\sigma_2W^A$.

Thus we are led to consider a theory with a Non-Abelian gauge
supermultiplet, in interaction with a number of chiral superfields.
Naive expectations is that the strong force will cause both gauginos and
matter fermions to form chiral condensates, and the matter condensates
will dynamically break supersymmetry. However the situation is not at
all that simple.

First of all, in the absence of matter, gaugino condensation occurs,
and, as expected, supersymmetry is not broken. Secondly, with chiral
matter, supersymmetry is not necessarily broken dynamically. If the
matter chiral multiplets have a common mass, supersymmetry is not broken
dynamically, even with strong coupling. Only when the matter is massless
can supersymmetry be broken dynamically, but then the lowest energy
configuration usually corresponds to infinite field values, except in
some very special, and more complicated models.

A very important tool in the study of dynamical breaking of {\bf global}
supersymmetry is the Witten index. We have seen, that because of the
supersymmetry algebra, it is easy to determine the breaking of
supersymmetry: supersymmetry is broken if and only if the state of
lowest energy is positive.

In supersymmetry theories, the potential is the sum of squares of F and
D terms, and it is positive definite. If the potential at minimum is
positive, it means that either the F and/or D terms have non-zero
values, and supersymmetry is broken.

Witten considers a supersymmetric theory in a finite volume V. To
preserve the translation symmetry he imposes periodic boundary
conditions, and examines the vacuum energy $E(V)$. He argues that if
$E(V)=0$ for finite $V$, it will remain so as the infinite volume limit
is taken.

The Hamiltonian will have discrete eigenvalues. Its spectrum is made up
of two types of states, boson and fermions. They are distinguished by
the value of the operator $$e^{i\pi J_z}\ ,$$ which has value $1$ on
bosons, and $-1$ of fermions. He observes that states of finite energy
always come in degenerate boson fermion pairs. This is a result of the
algebra, which states that the supersymmetry generator is the square
root of the Hamiltonian.

The situation is entirely different for the zero energy states, since
the application of the supersymmetry generator to any zero energy state
does not produce  another state, since the energy is zero. There need
not be the same number of bosonic and fermionic states of zero energy.
Let there be $n_b$ bosonic and $n_f$ fermionic states of zero energy.

Now let us assume that the system undergoes adiabatic changes, such as
changes of couplings, and other parameters. The occupation number of
states at a given energy level will change, and so will the energy
eigenvalues. However, to preserve supersymmetry, the states will migrate
in pairs from one level to the next. For instance, one positive energy
pair may migrate into the zero energy state. In that case, both $n_b$
and $n_f$ increase by one. Alternatively, two states in the zero energy
state may migrate into a state of positive energy. In this case, both
$n_b$ and $n_f$ decrease by one unit. However in both cases, the
difference
$$\Delta\equiv n_b-n_f$$
is left unchanged. It is very insensitive to most changes in the system,
and thus can be computed more easily, for instance in the perturbative
regime, where it is easier to calculate.

What use is this difference? Suppose it is different from zero. Then,
necessarily $n_b$ and/or $n_f$ is itself different from zero, indicating
that the zero energy state is occupied: supersymmetry is unbroken.

On the other hand, if $\Delta=0$, it may mean one of two thing: either
$n_b=n_f\ne 0$, in which case supersymmetry is unbroken, {\it or}
$n_b=n_f=0$, which indicates that there are no states of zero energy and
the breaking of supersymmetry.

The idea is to compute $\Delta$ for a value of parameters which lends
itself to calculability. If it is not zero, supersymmetry is not broken.
If it is zero, one cannot say anything.

The Witten index can be computed for a super pure Yang-Mills theory,
where it is found to be equal to the rank of the group plus one. Thus,
as expected, supersymmetry is not broken.

It can also be computed when massive chiral matter is added. Again, it
is found to be non-zero. However, when the mass of the chiral
superfields is taken to zero, computation of the Witten index ceases to
be trustworthy, because the potential no longer favors small field
values.

Before leaving this topic, we should mention that the situation is
thought to be quite different in the case of local supersymmetry. There,
the gaugino condensate is capable of breaking supersymmetry. It is a
favorite scenario of superenthusiasts to believe that this is what
happens in nature: a strong QCD-like force causes gauginos to condense.
This breaks supersymmetry. This strong force operates in the {\it hidden
sector}, a sector of the theory that is connected with ours only by the
universal force of gravity. This is technicolor in the hidden sector!
Thus supersymmetry breaking appears in our phenomenological theories
through a universal mechanism, given in terms of soft breaking
parameters. No model in which this actually happens has been formulated,
but it is sociologically true. Next, we discuss the effective soft
breaking of supersymmetry this phenomenon is believed to cause in our
sector.

Thus supersymmetry breaking is added to the $N=1$ standard model in the
form of soft terms. The incredible thing is that this simple hypothesis
triggers spontaneous breaking of the electroweak symmetry!

In order to appreciate the rationale behind such a picture, it is useful
to reason by analogy with low energy chiral symmetry, which although an
approximate symmetry of nature, has proved to be very important in the
analysis of low energy strong interactions.

Assume for a moment that the energy available to your machines is below
that of a pion (yes there was such a time!). Perhaps someone had
postulated back then that the chiral symmetry limit is an interesting
limit in which to study the Strong Interactions. In this picture, the
chiral symmetry is spontaneously broken, and the nucleons are
supplemented by massless pions. It also predicts, in the form of low
energy theorems, the couplings of pion to matter. But massless pions
have not been seen, and this sounds like a pretty weird thing to do.
Clearly chiral symmetry must be broken, but how?

Contrast the situation to the point of view I have conveyed in these
lectures: it makes sense to generalize the $N=0$ Standard Model to $N=1$
supersymmetry. This means the invention of squarks, sleptons, gauginos,
etc..., together with predictions of their couplings with ordinary
none of which have been seen. Clearly supersymmetry must be
broken, but how?

\noindent {\sl Chiral symmetry}: The pion is found, and it is light in terms of
strong interactions; it  means that the picture of approximate chiral
symmetry makes sense.

\noindent {\sl Supersymmetry}: a gluino is found, and the whole thing
makes sense.

\noindent {\sl Chiral symmetry}: there is more than one pion, and theorists
postulate
that chiral symmetry breaking appears in the form of soft terms, with
definite symmetry characteristics. These are in turns used to derive sum
rules relating the masses of the pseudoscalars.

\noindent {\sl Supersymmetry}: Theorists assume soft supersymmetry breaking
with
definite symmetry characteristics. A simple assumption is universality
of the breaking; it happens to be one of the most economical ways to
introduce the soft breaking. Theorists deduce sum rules. This means that
the form of the supersymmetry breaking may some day be determined by
experiment, but only after as many measurements as the number of soft
breaking parameters.

\noindent {\sl Chiral symmetry}: The meaning of the soft chiral symmetry
breaking is now easily understood in terms of the underlying theory:
QCD; it has the same quantum numbers as the quark mass terms in the
Lagrangian.

\noindent {\sl Supersymmetry}: one expects that it is experiment which will
eventually determine the form of the soft supersymmetry breaking
parameters. If the breaking is found to be universal, this will be a
strong indication for the hidden sector scenario. Need we say that it
occurs most naturally in the Heterotic String theory?

This analogy is not perfect. For one chiral symmetry is a global
symmetry, while we expect supersymmetry to be a local symmetry. This is
a crucial difference, but the role of effective soft breaking terms is
similar, in the sense that they are manifestations of deeper theory in
both cases, and are used for phenomenology in the same way.

In the universal breaking picture, all the squarks, sleptons, and Higgs
are given a common supersymmetry breaking mass $m_0$.
The three gauginos are also given masses $M_i$. They need not all be the
same, without extra assumptions. If the mother theory is grand unified,
then it is natural to take all three masses to be the same. This is also
true of some string theories. One also implements the soft breaking mass
term of the Higgs with a parameter $B$. Finally, terms of dimension
three appear as sparticle-sparticle-Higgs interactions of the same
symmetry character as the Yukawa couplings. All these parameters are
supposed to have values in the hundreds of GeV range, reflecting the
strength of supersymmetry breaking.

The soft breaking parameters appear as boundary
conditions in the renormalization group
equations that govern the running of the same parameters.
The scale at which they are specified is assumed
to be in the deep ultraviolet, near or at Planck scale.

Several remarkable things are seen to happen. First of all, the square
of the mass of the Higgs that couples to the up quarks, starting from
its ultraviolet value, is seen to become negative in the infrared.
Amazingly, the evolution equations are such that it is the Higgs that
becomes tachionic, indicating the spontaneous breaking of electroweak
symmetry. This is possible only because of the large value of the top
quark mass. I have no time to cover this beautiful development in these
introductory lectures, but you should be left with the appropriate sense
of awe.

Fortunately for you, there is a lot of work yet to be done. I believe
that the most important question of deep theoretical interest is
dynamical supersymmetry breaking. I look forward someday to hear that
one of you has actually solved its mechanism.

\vskip 1cm
I wish to thank Professors
J. Donoghue and K.T. Mahanthappa for their kind hospitality during my
stay at the 1994 TASI. I also wish to thank the students for their keen
interest and challenging questions. To some I apologize for restoring
the $\sigma_2$, but I could not cope with that many
indices.This work was supported in part by the United States Department of
Energy under Contract No. DEFG05-86-ER-40272.

I have not included references in the text, since the lectures are quite
introductory. Rather I draw your attention to several excellent
elementary books on the subject, as well as to reviews and reviews of
reviews. These are

\noindent J. Bagger and J. Wess, {\it Supersymmetry and Supergravity},
Princeton University Press, second edition (1993).

\noindent P. West {\it Introduction to Supersymmetry and Supergravity},
World Scientific, Singapore (1990).

\noindent {\it Supersymmetry and Supergravity}, a collection of Physics
Reports, edited by M. Jacob (1986).
\end